\documentclass[letterpaper, 10 pt, journal, twoside]{ieeetran}

\IEEEoverridecommandlockouts                              % This command is only
\usepackage{amssymb}                                                       % use the \thanks \usepackage{algorithm2e}
\usepackage[algo2e]{algorithm2e} 
\usepackage{ifpdf}
\usepackage{amsmath}
\usepackage{amsfonts}
\usepackage{mathrsfs}
\usepackage{color}
\usepackage{graphicx}
\usepackage{subfigure}
\usepackage{epstopdf}
\usepackage{lineno}
\usepackage{cite}
\usepackage{enumerate}
\usepackage{algorithm, algorithmic}
\usepackage{bm}

\newcommand{\cov}{{\rm{cov}}}

\newcommand{\Eb}{\mathbf{E}}

\newcommand{\RBB}{\mathbb{R}}

\newcommand{\AC}{\mathcal{A}}

\newcommand{\DC}{\mathcal{D}}

\newcommand{\HC}{\mathcal{H}}
\newcommand{\IC}{\mathcal{I}}
\newcommand{\KC}{\mathcal{K}}
\newcommand{\MC}{\mathcal{M}}
\newcommand{\NC}{\mathcal{N}}

\newcommand{\PC}{\mathcal{P}}

\newcommand{\RC}{\mathcal{R}}
\newcommand{\SC}{\mathcal{S}}
\newcommand{\TC}{\mathcal{T}}

\newcommand{\XC}{\mathcal{X}}
 
\newtheorem{theorem}{Theorem}
\newtheorem{definition}{Definition}

\newtheorem{lemma}{Lemma}     
\newtheorem{remark}{Remark}
\newtheorem{proposition}{Proposition}
\newtheorem{assumption}{Assumption}

\newcommand{\eq}{&\hspace{-0.5em}=\hspace{-0.5em}&} 
\newcommand{\qle}{&\hspace{-0.5em} \le\hspace{-0.5em}&}
\newcommand{\tr}{{\rm{tr}}}

\title{\LARGE \bf
Infinite-horizon optimal scheduling for feedback control}

\author{Siyi Wang and Sandra Hirche, \IEEEmembership{Fellow, IEEE}
\thanks{This work has been funded by the German Research Foundation (DFG) under the grant number 315177489 as part of the SPP 1914 (CPN), and by the 
Federal Ministry of Education and Research
of Germany in the programme of “Souverän. Digital. Vernetzt. ” under the
joint project 6G-life (Project ID: 16KISK002).}
\thanks{Siyi~Wang and Sandra~Hirche are with the Chair of Information-oriented Control (ITR), Technical University of Munich, Germany, e-mail: \{siyi.wang, hirche\}@tum.de.}}

\graphicspath{{figures/}}

\begin{document}

\maketitle
\thispagestyle{empty}
\pagestyle{empty}

\begin{abstract}
Emerging cyber-physical systems impel the development of communication protocols that optimize resource utilization. This article investigates infinite-horizon optimal scheduling for resource-aware networked control systems by addressing the rate-regulation tradeoff. Consider a scenario where the sensor and the controller communicate via a networked channel, the transmission scheduling problem is formulated as a Markov decision process on unbounded general state space controlled by scheduling decisions. The value of information (VoI) serves as a metric to assess the importance of sensory data for transmission. We derive the optimal scheduling law for feedback control based on VoI and show that it is deterministic and stationary, with an explicit expression obtained via value iteration. The closed-loop system under the designed scheduling law is shown to be stochastically stable. By analyzing the dynamic behavior of the iteration process, we show that the VoI function and the optimal scheduling law exhibit symmetry. Furthermore, when the system matrix is diagonalizable, the VoI function is monotone and quasi-convex. Consequently, the optimal scheduling law is shown to exhibit a threshold structure and takes a quadratic form, with the threshold region explicitly characterized. Finally, the numerical simulation illustrates the theoretical result of the VoI-based scheduling. 
\end{abstract}


\section{Introduction}\label{sec:introduction}
Networked control systems (NCSs) are control systems where feedback control loops are closed over a communication network \cite{walsh2001scheduling}. 
Application domains include fields such as robotics \cite{barfoot2017state},  smart energy grids \cite{singh2014stability},  and autonomous driving
\cite{li2017dynamical}, etc.
Traditional networked control designs often assume that sensory data are accessed periodically to update control inputs. However, due to communication resource constraints and channel bandwidth limitations, not all data are valuable enough to warrant transmission. Efforts have been made to enhance communication resource utilization efficiency, including the development of event-triggering mechanisms \cite{lunze2010state,wang2010event} and the optimization of communication protocols tailored to specific applications \cite{soleymani2022value}, etc.

The scheduling law determines whether sensory data should be transmitted, thereby regulating the system's behavior. Since control quality and resource utilization efficiency are inherently interrelated, it is essential to coordinate control laws with scheduling mechanisms to achieve a desired control objective.  In stochastic systems, the dual effect of control \cite{aastrom2012introduction} means that the controller simultaneously: i) influences state evolution and ii) probes the system to reduce state uncertainty. However, the effects of control and scheduling are decoupled when the scheduling criterion is independent of past control actions  \cite{ramesh2011on,molin2013on}, known as the separation principle in event-triggered stochastic control \cite{aastrom2012introduction}.
Building on these findings, this study fixes the control law as a certainty equivalence controller \cite{aastrom2012introduction} derived from the corresponding deterministic optimization problem and focuses on the optimal scheduling design. 
To evaluate the semantic importance of the data, we employ the concept of the VoI for feedback control \cite{soleymani2021value}, which 
quantifies the uncertainty of decision makers regarding control task achievement.

Additionally, studies on optimal state-based scheduling mainly focus on finite-horizon performance, aiming to minimize costs at each individual instance \cite{soleymani2021value, molin2014suboptimal}. However, the problem's dimensionality grows exponentially with the horizon length, resulting in high computational complexity.
This article addresses the expected average cost minimization over the infinite horizon. It is meaningful because i) the optimal policies solved from the average-cost minimization are typically stationary and ii) it offers insightful analysis. 
The classic work \cite{bertsekas2005dynamic} analyzes the average-cost minimization problem on a countable state space.
% Some properties, such as convergence and monotonicity of the optimal action law, are analyzed.  
Additionally, extensive research has explored average-cost problems on Borel state spaces, including scenarios with unbounded per-stage costs and compact action spaces \cite{hernandez1990average}, as well as cases with non-compact action spaces \cite{gordienko1995average}.

\subsection{Related works}
There are some works addressing optimal scheduling for remote estimation, see \cite{lipsa2011remote,molin2017event,rabi2008optimal,leong2016sensor,leong2018transmission,duan2022sensor,wang2023value}. 
For example, \cite{lipsa2011remote} studies the optimal event-triggered estimation for a first-order system and shows that a threshold-based pre-processor and a
Kalman-like filter at the estimator are jointly optimal. Duan et~al. \cite{duan2022sensor} investigate sensor scheduling for distributed estimation and show that the optimal policies have a threshold structure.  
Some works, e.g., \cite{leong2016sensor,leong2018transmission}, address the optimal event-triggered estimation are variance-based. However, state-based scheduling policies outperform variance-based ones by leveraging real-time sensory information
\cite{wu2013can}.
When addressing the optimal event triggering for control, the problem becomes complex due to the dual effect of the controller. In this regard, \cite{molin2013on} shows that the separation principle holds in stochastic optimization problems with resource constraints and identifies an optimal solution pair where the triggering depends only on primitive random variables.  
Additionally, \cite{soleymani2021value} develops a systematic framework to jointly co-design the control and communication, where a globally optimal policy profile consists of a symmetric triggering policy, and a certainty equivalence control law \cite{soleymani2022value}. However, all the above works address optimal scheduling for NCSs over a finite horizon. To mitigate the computation burden, \cite{molin2019scheduling,wang2021value} employ rollout algorithms to approximate the cost-to-go, yielding a suboptimal solution. In this article, we study the optimal state-based scheduling for feedback control addressing long-term performance. 
Another related work is \cite{xu2004optimal}, which assumes bounded state space and bounded per-stage cost. In this study, we generalize it to unbounded state space and unbounded stage cost. 

One main novelty of this work lies in analyzing the structure of the optimal state-based scheduling law. Existing literature,  e.g., \cite{lipsa2011remote,ren2017infinite,chakravorty2018sufficient}, provide relevant insights. For instance, \cite{lipsa2011remote,chakravorty2018sufficient} investigate infinite-horizon optimal scheduling problems on real space or finite integer set and show that the value function solved from the Bellman equation is symmetric, monotone, and quasi-convex. Additionally, \cite{ren2017infinite} investigates the optimal co-design of transmission power scheduling and remote estimation and shows that the power scheduling is monotone when the system matrix is a scalar or an orthogonal matrix. All these works use the majorization technique to prove that Bellman operation preserves the symmetry and monotonicity of value functions under certain conditions. 
In this work, we develop a novel framework to analyze the structure of the optimal scheduling law when the system matrix is diagonalizable.

\subsection{Our contributions}
In this article, we investigate the optimal scheduling for resource-aware NCSs by addressing long-term rate-regulation tradeoff. 
First, by leveraging the result of \cite{ramesh2011on,molin2013on}, we fix the control law as a certainty equivalence controller. This ensures that the scheduling design is unaffected by the dual effect of control. The optimal scheduling problem is formulated as a Markov decision process (MDP) on a Borel state space controlled by scheduling decisions, which is solved using an average-cost minimization approach  \cite{hernandez1990average,montes1994average}.

We summarize our main contributions as follows: 
1) by verifying specific conditions, we show that the formulated average-cost MDP 
admits an optimal solution triplet, which includes the optimal stationary scheduling law, the optimal average cost, and the differential cost function (Sec.~\ref{sec:average-cost MDP}); 
2) we use value iteration algorithm to obtain the differential cost and show that the iteration sequence converges. Furthermore, we solve the Bellman equation on a truncated space and show the relation between this truncated solution and the original solution. (Sec.~\ref{sec:value iteration});
3) we quantify the VoI metric and derive the VoI-based optimal stationary scheduling law. The error dynamics under the VoI-based scheduling law is shown to be bounded, which guarantees the stochastic stability of the closed-loop system (Sec.~\ref{sec:optimal scheduling}); 
4) we analyze the dynamic behavior of the iterative algorithm and show that the differential cost, its expectation, and the VoI function are symmetric. Additionally, when the system matrix is diagonalizable, the VoI function is monotone and quasi-convex. Based on these, the VoI-based scheduling law is shown to be of threshold type and takes a quadratic form. This finding simplifies the computation and implementation of the optimal scheduling policy (Sec.~\ref{sec:threshold structure}).

\subsection{Outlines}
The remainder of this article is structured as follows: Section \ref{sec:preliminaries} introduces system model and problem formulation. Section \ref{sec:main result} presents the main result on the optimal stationary scheduling design and analysis. Section \ref{sec:simulation} uses numerical simulation to illustrate theoretical results. Section \ref{sec:conclusion} concludes this work.

\noindent\textbf{Notations:} For a symmetric positive definite matrix $\Sigma$, let $\Sigma^{\frac{1}{2}}$ denote the Cholesky decomposition of $\Sigma$ such that $\Sigma  = (\Sigma^{\frac{1}{2}})^{\top}\Sigma^{\frac{1}{2}}$. Let $\{w_{k}\} $ denote the sequence of random process $w_{k}$. Let $z_{[0:k]}$ denote the state sequence $\{z_0, z_1, \dots, z_k\}$. 
Let $ x \sim \NC(\mu, \Sigma)$ denote that the random variable $x$ follows a normal distribution with mean $\mu$ and covariance $\Sigma$. Let $\Eb[\cdot]$, $\Eb[\cdot\mid\cdot]$ and $\cov[\cdot]$ denote the expectation, the conditional expectation and the covariance of the random variable, respectively. Let $\mathbf{0}_m$ denote a $m$-dimension vector with all entries being $0$. Unless otherwise noted, we use $\mathbf{0}$ to denote a $n$-dimension vector with all entries being $0$. Let $(\mathbb{R}^n,\mathscr{B})$ be a measurable space and $\mu$ be a signed measure defined on the sigma-algebra $\mathscr{B}$. The total variation distance of two probability measures $\mu$ and $\nu$ is defined as $\|\mu -\nu\|_{tv}: = 2\sup_{\mathcal{B}\in \mathscr{B}} |\mu(\mathcal{B})-\nu(\mathcal{B})|$. For matrix $A$, let $\bar{\rho}(A)$ denote the maximum eigenvalue of $A$.  

\section{Preliminaries}\label{sec:preliminaries}
We consider a resource-constrained feedback control system closed over a communication network, as in Fig.~\ref{fig:system architecture}. The local event-based scheduler determines whether to send out Kalman estimate. The controller generates control signals based on a remote estimator. 
 
\subsection{System model}
Consider the discrete-time stochastic dynamical system: 
\begin{eqnarray}\label{eq:plant}
x_{k+1}  \eq A x_{k}  +B u_{k}  +w_{k}    \nonumber \\  
y_{k}  \eq C x_{k} + v_{k}, 
\end{eqnarray}
where  $x_k \in \mathbb{R}^{n}$, $u_{k}\in \mathbb{R}^{m}$, $y_k \in \mathbb{R}^{p}$ are the state vector, the control force and the measurement, respectively. The system matrices are $A\in \mathbb{R}^{n\times n}$, $B \in \mathbb{R}^{n\times m}$, $C\in \mathbb{R}^{p\times n}$, where  the pair $(A,B)$ is controllable and  $(A,C)$ is observable. The process noise $w_k \in \RBB^{n} \sim \NC(0, W)$ and the measurement noise $v_{k}\in \RBB^{p} \sim \NC(0, V)$ are assumed to be independent identically distributed (i.i.d.)  Gaussian processes with zero mean and positive semidefinite variances $W$ and $V$, respectively. 
The initial state  $x_0 \sim \NC(\bar{x}_{0}, R_{0}) $  is a random vector with mean $\bar{x}_{0}$ and positive semi-definite covariance $R_{0}$, which is statistically uncorrelated with  $w_k$, $v_k$ for all $k$. 
\subsection{Network model}
\subsubsection{Local sender (Kalman filter and scheduler)}
\begin{figure}[t]
    \centering
\includegraphics[width=0.4\textwidth]{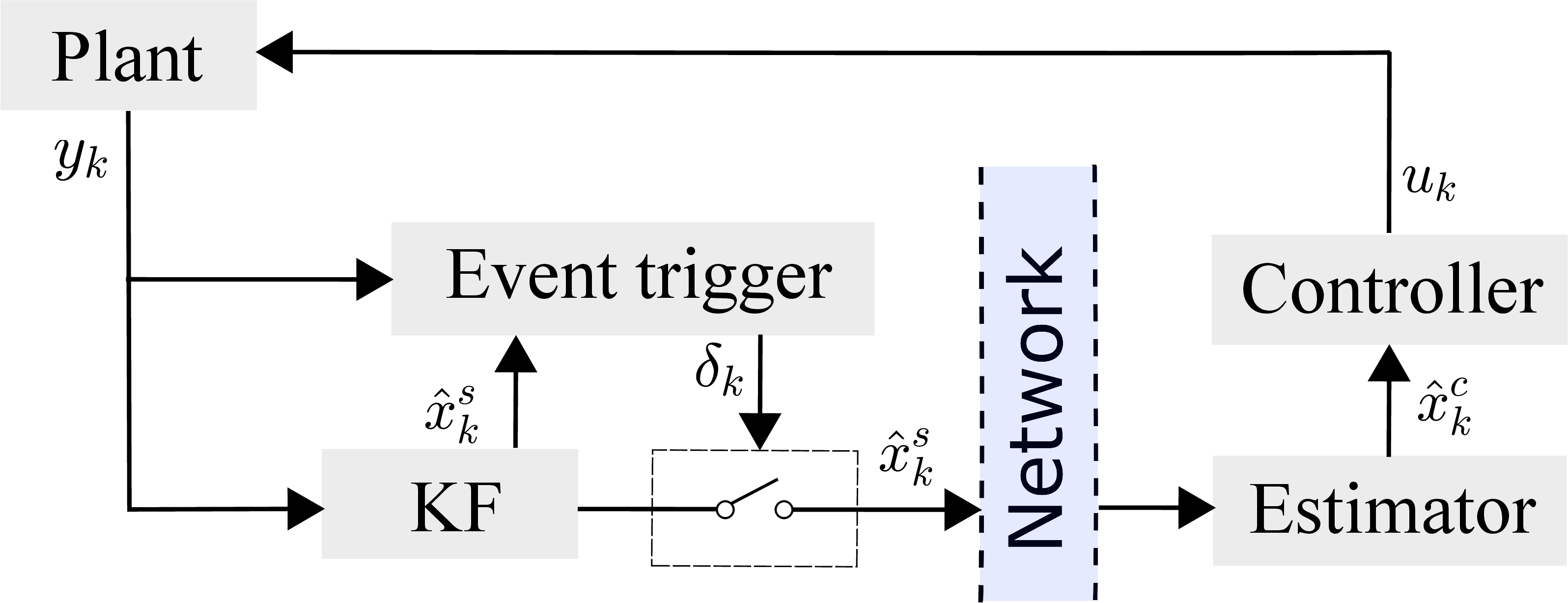}
    \centering 
    \caption{NCS architecture}
    \label{fig:system architecture}
\end{figure}

As in Fig.~\ref{fig:system architecture}, the local sensor accesses the process measurement periodically, and the  Kalman filter computes the local estimate: 
\begin{eqnarray}\label{eq:Kalman filter}
    \hat{x}_{k+1}^s \eq \Eb[x_{k+1} \vert \IC_{k+1}^{s}] \nonumber \\
    \eq A\hat{x}_k^s  + Bu_k    + K_{k} \big(y_{k+1} - C(A\hat{x}^{s}_k + Bu_k ) \big),
\end{eqnarray}
where  $\hat{x}_0^s = \bar{x}_0$ is the initial value,  $K_{k} \in \RBB^{n\times p}$ denotes the Kalman gain matrix,  and $\IC_k^s$ denotes the local information set by time $k$.  Denote the local estimation error as $\hat{e}_k^s = x_k -  \hat{x}_k^s$ and denote its covariance as $P_k^s=\cov[\hat{e}_k^s]$.  Since the pair $(A,C)$ is observable, the estimation error covariance $P_k^s$ and the Kalman gain $K_k$ converge to the steady values $P^s$ and $K = P^sC^\top(CP^sC^\top + R)^{-1}$, where $P^s$ satisfies the algebraic Riccati equation $    P^s = A(P^s-P^sC^\top(CP^sC^\top + V)^{-1}CP^s)A^\top + W$.

Generally, sending the estimate enables the remote estimator encode more information compared to sending the measurement \cite{xu2005estimation}. The event trigger decides whether to send out the Kalman estimate $\hat{x}^{s}_{k}$ through the communication network.  Denote $\Gamma$ and $\AC:= \{0,1\}$ as the admissible scheduling law and the scheduling decision set, respectively. The  transmission decision $\delta_{k} \in \AC$ is determined by
\begin{eqnarray}\label{eq:triggering law definition}
   \delta_{k} = \gamma(\IC_{k}^{s}) =
   \left\{ 
    \begin{array}{ll}
        1 &   {\rm{transmission~occurs}}\\
         0& {\rm{otherwise}},
    \end{array}\right.
\end{eqnarray}
where  $\gamma \in \Gamma$ denotes the scheduling law, and $\IC_{k}^{s}= \{ y_{0:k},  u_{0:k-1},\delta_{0:k-1} \}$ with the initial value $\IC_{0}^{s} = \{y_{0} \}$.

\subsubsection{Remote receiver (linear estimator and controller)} 
Assume that the communication network induces a one-step delay. The state received by the remote side is
\begin{eqnarray*}
z_{k} = \left\{ \begin{array}{ll}
\hat{x}_{k-1}^{s}     &  {\rm if}~ \delta_{k-1} = 1\\
 \emptyset    & {\rm otherwise}.
\end{array}\right.
\end{eqnarray*}
Then a linear estimator computes the system state by time $k$:
\begin{eqnarray}\label{eq:remote estimator}
\hat{x}^{c}_{k} \eq  \Eb[x_{k} \vert \IC_{k}^{c}]  =   
A\Eb[x_{k-1}| \IC_{k}^c]+B u_{k-1} 
\end{eqnarray}
with the initial value $\hat{x}^{c}_{0} = \bar{x}_0$, and   $\IC_{k}^{c} = \{z_{0:k}, u_{0:k-1}, \delta_{0:k-1}\} $ is the remote information set. The control law is given by 
\begin{eqnarray}\label{eq:controller}
u_{k} = f ( \IC_{k}^{c})= -L  \Eb[x_{k} \vert \IC_{k}^{c}],
\end{eqnarray} 
where the control gain $L$ will be designed later.
To decouple the effects caused by the control and the scheduling decisions, we fix the control law as the certainty equivalence controller.
Moreover, given the one-step delay assumption, the information pattern between the controller and the scheduler becomes nested, and thus the separation principle holds \cite{molin2014suboptimal}.

\subsection{Problem statement} 
In this article, we aim to design an optimal scheduling law to minimize the long-term  regulation and communication costs:  
\begin{eqnarray} \label{eq:original opt}  
     \min_{  \gamma \in \Gamma  }  & \Psi(\gamma) =  V(\gamma)  + \RC(\gamma),
\end{eqnarray}
where $V( \gamma)$ denotes the regulation cost, which is measured by the infinite-horizon LQG function: 
% $J = \limsup_{T \rightarrow \infty} \frac{1}{T}\Eb\big[ \sum_{k=0}^{T-1 }   x_{k}^\top Qx_{k} + u_{k}^\top R u_{k}  \big]$, 
\begin{eqnarray}\label{eq:LQG}
V(\gamma) = \lim_{T \rightarrow \infty} \frac{1}{T}\Eb\bigg[ \sum_{k=0}^{T }  \hspace{-1em}& x_{k}^\top Qx_{k} + u_{k}^\top R u_{k}  \bigg],  
\end{eqnarray} 
where the matrices $Q$ and $R$ are positive 
 semidefinite and positive definite, respectively. Moreover, the pair $(A, Q^{\frac{1}{2}})$ is assumed to be detectable, with $Q  = (Q^{\frac{1}{2}})^\top Q^{\frac{1}{2}}$.  
The communication cost is  measured by $$\RC = \lim_{T \rightarrow \infty} \frac{1}{T}\Eb\big[\sum_{k=0}^{T}\theta \delta_{k} \big],$$ 
where $\theta $ denotes the unit transmission price.  When $\theta$ represents a tradeoff multiplier, the objective function $\Psi(\gamma)$ is interpreted as a tradeoff between the regulation cost and the communication rate.  
% Note that the average cost minimization problem usually assume a stationary system, e.g., the cost per state and random disturbance statistics do not change over time, which is often satisfied in practice. 

\section{Main results}\label{sec:main result}

Before presenting the main result, we introduce the following lemmas. These lemmas derive the certainty equivalence controller by solving a corresponding deterministic optimization problem \cite{aastrom2012introduction} and simplify the problem \eqref{eq:original opt} based on the chosen controller.
\begin{lemma}\label{lemma:controller}
\cite{aastrom2012introduction}
Consider the LQG function \eqref{eq:LQG}, the optimal control law minimizing \eqref{eq:LQG} is the certainty equivalence controller \eqref{eq:controller} with the control gain 
\begin{eqnarray}\label{eq:control gain}
    L = -(R  + B^{\top}S B)^{-1}B^{\top}SA,
\end{eqnarray} 
where $S$ is the solution of the algebraic Riccati equation:
\begin{eqnarray}\label{eq:Riccati matrix} 
S  =  Q+A^{\top}\big(S-SB(R  + B^{\top}S B)^{-1}B^{\top}S\big)A. 
\end{eqnarray}  
\end{lemma}
               
Applying the algebraic Riccati equation \eqref{eq:Riccati matrix} recursively to \eqref{eq:original opt}, we obtain 
\begin{eqnarray}\label{eq:opt 1}
    \Psi (\gamma ) \eq \lim_{T\rightarrow \infty}\frac{1}{T}  \Eb\bigg[x_{0}^{\top}Sx_{0} + \sum_{k=0}^{T} \big( w_{k}^{\top}S w_{k} \nonumber \\ 
    && \hspace{-2.7em}+ (u_{k}+L x_{k})^{\top} (R  + B^{\top}S B) (u_{k}+L x_{k}) + \theta \delta_{k} \big)\bigg]. 
\end{eqnarray}  
Denote the remote estimation error as  $\hat{e}^{c}_{k} = x_{k} - \hat{x}_k^c$.  
Denote the mismatch between the local estimate \eqref{eq:Kalman filter} and the remote estimate \eqref{eq:remote estimator} as 
\begin{eqnarray}\label{eq:estimate mismatch}
    e_{k}=   \hat{x}^{c}_{k} - \hat{x}_{k}^{s}. 
\end{eqnarray}
\begin{lemma} 
Under the certainty
equivalence controller \eqref{eq:controller}, the optimization problem \eqref{eq:original opt} is reduced to 
\begin{eqnarray}\label{eq:simplified opt}
    \min_{\gamma \in \Gamma} J(\gamma),  
\end{eqnarray}
where
\begin{eqnarray}\label{eq:Vk}
J(\gamma) = \lim_{T \rightarrow \infty}\frac{1}{T } \sum_{k=0}^{T}\Eb\big[ g(e_{k},\delta_{k}) \big| \IC_k^s\big] 
\end{eqnarray}
with the per-stage cost 
\begin{eqnarray}\label{eq:stage cost}
g(e_{k} ,\delta_{k}) = \theta \delta_{k} + (1-\delta_{k}) e_{k}^{\top}A^\top \Sigma Ae_{k}
\end{eqnarray}  
with $\Sigma = A^\top SB(R  + B^{\top}S B)^{-1}B^\top S A $ and 
\begin{eqnarray}\label{eq:estimation mismatch}
e_{k+1} = (1-\delta_k) Ae_k + \xi_k 
\end{eqnarray}
with  
$ \xi_{k} = KC(A\hat{e}^{s}_{k} +w_{k}) + Kv_{k+1} \in \RBB^{n}$.
\end{lemma} 
\textbf{Proof}. It follows the proof of \cite{soleymani2021value,soleymani2022value}. Though \cite{soleymani2021value,soleymani2022value} investigate the finite-horizon case, its extension to the infinite-horizon case is straightforward. 
From the remote estimator \eqref{eq:remote estimator}, we have 
\begin{eqnarray} \label{eq:remote estimation}
\hat{x}_{k+1}^c \eq A \hat{x}_k^c + Bu_k \nonumber \\ 
&& +\delta_kAe_k + (1-\delta_k)A\Eb[\hat{e}_k^c|\IC_k^c,\delta_k=0],
  % \hat{e}^{c}_{k+1} \eq  \left\{ \begin{array}{ll}
  % A\hat{e}^{s}_{k} + w_{k}     & {\rm if}~\delta_{k}=1 \\
  % A\hat{e}^{c}_{k} + w_{k}       & {\rm otherwise}.
  % \end{array}\right. 
\end{eqnarray}
where the last term is inferred from non-triggering. Under the control law $f$ that is symmetric with respect to innovation $y_k - C\Eb[x_k|\IC_{k-1}^s]$, we have 
$\Eb[\hat{e}_k^c|\IC_k^c,\delta_k=0] = 0$, see \cite{soleymani2021value}. Thus, the remote estimation error evolves as 
\begin{eqnarray}\label{eq:remote estimation error}
    \hat{e}_{k+1}^c = A\hat{e}_k^c - \delta_k Ae_k + w_k, 
\end{eqnarray}
where $e_k$ evolves as \eqref{eq:estimation mismatch}, which is from substituting \eqref{eq:Kalman filter} and \eqref{eq:remote estimation} into \eqref{eq:estimate mismatch}.  Substituting the certainty equivalence controller \eqref{eq:controller} into \eqref{eq:opt 1}, we obtain 
\begin{eqnarray}\label{eq:simplified cost 1}
\Psi (\gamma ) \hspace{-0.2em}= \hspace{-0.2em}   \tr(SW) \hspace{-0.2em}+\hspace{-0.2em} \lim_{T \rightarrow \infty}\frac{1}{T } \Eb\bigg[\sum_{k=0}^{T} (\hat{e}^{c}_{k+1})^{\top}\Sigma \hat{e}^{c}_{k+1}  \hspace{-0.2em} + \hspace{-0.2em}\theta\delta_k \bigg| \IC_k^s\bigg].  
\end{eqnarray}
% \begin{eqnarray}\label{eq:conditional iota}  
% &&\hspace{-2.5em} \Eb\big[(u_{k}+L x_{k})^{\top} \Theta (u_{k}+L x_{k})\big] \nonumber \\
% &&\hspace{-3.5em}=\Eb\Big[\big(u_k+L\Eb[x_k|\IC_k^c]\big)^\top \Theta\big(u_k+L\Eb[x_k|\IC_k^c]\big) \big| \IC_{k-1}^s \Big]  \nonumber \\   
% && \hspace{-2.5em}+ 2\Eb[(u_k+L\Eb[x_k|\IC_k^c])^\top \Theta L\hat{e}_k^c| \IC_{k-1}^s] \nonumber \\ 
% &&\hspace{-2.5em} +   \Eb[(L\hat{e}_k^c)^\top \Theta L\hat{e}_k^c| \IC_{k-1}^s]   
% \end{eqnarray}
% which follows from the tower property of the conditional expectation and the fact that the $\sigma$-algebra generated by $\IC_{k-1}^s$ is a subset of the $\sigma$-algebra generated by $\IC_k^c$, i.e., $\sigma( \IC_{k-1}^s) \subseteq \sigma(\IC_k^c)$. In the right-hand side of \eqref{eq:conditional iota}, the first term becomes zero by \eqref{eq:controller}, and the second term becomes zero since $ 
%  \Eb[(u_k+L\Eb[x_k|\IC_k^c])^\top \Theta L\hat{e}_k^c| \IC_{k-1}^s]   =   \Eb[(u_k+L\Eb[x_k|\IC_k^c])^\top \Theta L\Eb[\hat{e}_k^c|\IC_k^c] \IC_{k-1}^s]=0$.
Moreover, by \eqref{eq:remote estimation error}, we have
\begin{eqnarray}\label{eq:error cost funcion}
\hspace{-2em}&&\hspace{-0.5em}\Eb[   (\hat{e}^{c}_{k+1})^{\top}\Sigma \hat{e}^{c}_{k+1}   \vert \IC_{k}^{s} ] \nonumber\\ 
% \hspace{-1em}\eq \hspace{-0.5em}\Eb[    (A\hat{e}^{c}_k)^{\top}\Sigma A\hat{e}^{c}_k + \delta_k^2(Ae_k)^\top \Sigma Ae_k + w_k^\top \Sigma w_k \nonumber \\
% \hspace{-1em}  &&\hspace{-0.5em}- 2\delta_k(Ae_k)^\top \Sigma A\hat{e}_k^c - 2\delta_k(Ae_k)^\top \Sigma w_k + 2 (A\hat{e}_k^c)^\top \Sigma w_k \vert \IC_{k}^{s} ] \nonumber \\
\hspace{-2em}\eq \hspace{-0.5em}\Eb\big[(1-\delta_{k})  e_{k}^{\top}A^{\top}\Sigma Ae_{k}\vert \IC_{k}^{s}\big]  \hspace{-0.2em}+\hspace{-0.2em} \tr(A^{\top} \Sigma AP^{s}) \hspace{-0.2em}+\hspace{-0.2em} \tr(\Sigma W),
\end{eqnarray} 
of which derivation see \cite{soleymani2022value}. Substituting  \eqref{eq:error cost funcion} into \eqref{eq:simplified cost 1}, we obtain \eqref{eq:Vk}. 
\hfill $\blacksquare$

Since the separation principle holds, the optimal scheduling law that will be designed later, alongside the certainty equivalence controller \eqref{eq:controller}, remains optimal in the joint design of control and communication for the rate-regulation tradeoff \eqref{eq:original opt}. This structural result has been studied in \cite{molin2010structural,soleymani2022value}, and is applicable to the infinite-horizon scenarios. The following sections focus on the optimal scheduling design. 

\subsection{Average-cost MDP}\label{sec:average-cost MDP}
In this section, we formulate \eqref{eq:simplified opt} as an average-cost optimization problem and obtain its stationary solution by solving the Bellman optimality equation. The infinite-horizon optimization problem \eqref{eq:simplified opt} is formulated as a MDP with the control model $(\XC,\AC,\PC,g)$ \cite{hernandez2012discrete}, where $\XC:=\mathbb{R}^n$ is Borel 
 state space,  $\AC$ is the action space, $\PC$ is the Borel measurable transition kernel defined on $(\XC,\AC)$ and  $g$ is the stage cost given by \eqref{eq:stage cost}. 
We use $\AC(e)$ to denote that for each $e \in \XC$, a nonempty set $\AC(e)$ is associated, and its elements $\delta \in \AC(e) $ are feasible actions. 
The stochastic kernel $\PC$ on $(e,\delta) \in \XC \times \AC$ is defined according to \eqref{eq:estimation mismatch}. Since the random variable $\xi_k$ is composed of $w_{0:k}$ and $v_{0:k+1}$, we obtain that $\xi_k$ follows a distribution of $\NC(\mathbf{0},\Xi)$  with $\Xi:= K\big( C(A^{\top}P^sA + W)C^{\top} +V \big)K^{\top}$ \cite{soleymani2022value}. Then, the stochastic kernel $\PC$ on $(e,\delta) \in \XC \times \AC$  is given as
\begin{eqnarray}\label{eq:transition}
\PC\big({\rm d} y \vert e , \delta \big):\hspace{-0.5em}\eq p\big(y\vert e,\delta\big){\rm d}y \nonumber \\
\eq \big[ \delta p_{\xi}(y)   + (1-\delta ) p_{\xi}(y-Ae )\big] {\rm d}y,  \label{eq:decision based pdf}
\end{eqnarray}
where $p$ is the transition probability function from the current state $(e,\delta):= (e_k, \delta_k)$ to the next state $y: = e_{k+1}$, and
$p_{\xi}$ is the probability density function of the random variable $\xi \sim \mathcal{N}(\mathbf{0},\Xi)$, expressed as 
$ p_{\xi}(y) =  \frac{1}{(2 \pi)^{\frac{n}{2}} \vert \Xi  \vert^{\frac{1}{2}}} \exp (-\frac{y^{\top}\Xi^{-1}y}{2}).$

The following proposition presents the Bellman optimality equation with the infinite-horizon performance criterion, whose solution is equivalent to that of the original optimization problem \eqref{eq:simplified opt}. For more explanation regarding the average-cost MDP, see Remark~\ref{remark:MDP}. 
\begin{proposition}\label{proposition:bellman}
Consider the optimization problem \eqref{eq:simplified opt}, there exists a constant $j^\ast$, a continuous function $h: \XC \rightarrow \mathbb{R}$ such that
\begin{enumerate}
\item the pair $(j^\ast, h)$ is a solution to the average-cost optimality equation:
\begin{align}\label{eq:bellman}
\hspace{-1em}  j^\ast+  h(e) &= \TC h(e) \nonumber \\ 
  &= \min_{\delta \in \AC(e) } \left\{g(e,\delta) + \int h(y) \PC({\rm d} y \vert e,\delta) \right\}, 
\end{align}
for $\forall~e\in \XC$, where $ \TC$ is the Bellman operator. 
\item  there exists a stationary policy $\gamma^\ast \in \Gamma $ such that $\gamma^\ast(e) \in \Gamma(e) $ minimizes the right-hand side of \eqref{eq:bellman}, where $\gamma^\ast$ is average-cost optimal, and $J^\ast = \inf_{\gamma\in\Gamma} J_\gamma(e) = j^\ast$, for $e \in \XC$. 
    \end{enumerate}
\end{proposition}
\textit{Proof}.
We use a 'vanishing discount' approach \cite{hernandez1990average}, which analyzes associated discounted cost problems with a varying discount factor $\alpha \in (0,1)$,  to derive the average-cost optimal solution. Denote the corresponding discounted cost under policy $\gamma$ as $   J_\alpha^\gamma(e):= \sum_{k=0}^{\infty} \alpha^k \Eb[g(e_k,\gamma(e_k)) \vert e_0=e]$,   
of which minimum is defined as $J_{\alpha}^\ast(e) = \inf_{\gamma \in \Gamma}J_\alpha^\gamma(e)$.   
Let $\alpha$ approach $1$, then the discounted cost problem approaches the average-cost minimization problem under the following conditions \cite{hernandez2012discrete}. In the following, we omit the subscript when $\alpha=1$, e.g., denote $J_1^\ast$  as $J^\ast$. 
According to \cite{montes1994average,hernandez2012discrete}, we need to verify the following conditions: 
\begin{enumerate}
    \item $g(e,\delta)$ is nonnegative, lower semicontinuous (l.s.c.) and inf-compact on Borel set $\KC = \{(e,\delta)\vert e \in \XC, \delta \in \AC \}$. 
    \item The function $\int h(y)\PC({\rm d}y \vert e,\delta)$ is l.s.c. and bounded in $(e,\delta) \in \KC$, whenever $h$ is l.s.c. and bounded in $e$. 
    \item The multifunction $\gamma(e): \XC \rightarrow \AC(e)  $ is upper semicontinuous. 
        \item There exists a nonnegative constant $M$,  and $\alpha_0 \in (0, 1)$ such that $J_{\alpha}^\ast(e) < \infty$ for every $e$ and $\alpha \in (0,1)$, and there exists a  fixed state $\varepsilon$ such that $(1-\alpha)J_{\alpha}^\ast(\varepsilon) \le M$ for $\forall~\alpha \in [\alpha_0,1)$. 
    \item Define $h_{\alpha}(e): = J_{\alpha}^\ast(e)-J_{\alpha}^\ast(\varepsilon)$ for all $e$. There exist nonnegative constants $N \in \mathbb{R}$ and a nonnegative measurable function $b(\cdot): \XC \rightarrow \mathbb{R}$, such that 
    $-N \le h_\alpha(e) \le b(e)$, for every $e \in \XC $ and $\alpha \in [\alpha_0,1)$, where the state $\varepsilon$ is the same as 4).  Moreover, $\int b(y)\PC({\rm d}y\vert e, \delta) < \infty$ for every $\delta \in \AC$ and $y,e \in \XC$.
    \item Under conditions 4-5), there exists a constant $j^\ast$,  and a sequence of discount factors $\alpha_n \uparrow 1$ such that $\lim_{n\rightarrow \infty}(1-\alpha_n)J_{\alpha_n}^\ast(e) = j^\ast $ for $\forall e \in \XC$. Fix the $\{\alpha_n\}$, define $h_{\alpha_n}(e):= J_{\alpha_n}^\ast(e)-J_{\alpha_n}^\ast(\varepsilon)$, we have $\{h_{\alpha_n} \}$ is equicontinuous. 
\end{enumerate}
For condition 1), the function $g(e,\delta)$ is nonnegative and l.s.c. by its definition, as in \eqref{eq:stage cost}. As the components of $\AC$ are finite integers, the set $\{ \delta \in \AC \vert g(e,\delta) \le r \}$ is compact for every $e \in\XC$ and $r \in \mathbb{R}$.  Therefore, the $g(e,\delta)$ is inf-compact on $\KC$. 

For condition 2),
since $\PC({\rm d}y \vert e,\delta)$ 
is a continuous function on $(e,\delta)$, and the composition of two continuous functions is still continuous. Thus, we have $g(e,\delta)$ is continuous on $\KC$. Moreover, $\int h(y)\PC({\rm d}y \vert e,\delta)$ is bounded in $(e,\delta) \in \KC$ if $h(y)$ is bounded. Specifically, Assumption (b.2) of \cite{hernandez1990average} demonstrates that the system equation \eqref{eq:estimation mismatch} with state space $\mathbb{R}^n$  satisfies condition 2). 
Condition 3) holds as $\KC$ is convex, see Lemma~3.2 of \cite{hernandez1995numerical}. 
For condition 4), we set a decision sequence as $\bar{\gamma} = \{1,1,\dots,\}$, and obtain $J_{\alpha}^{\bar{\gamma}}(e) = \Eb[\sum_{k=0}^{\infty
   }   \alpha^k  g(e_k,1)\vert e_0=e] = \frac{\theta}{1-\alpha}.$
Thus, $    J_{\alpha}^\ast (e) \le J_{\alpha}^{\bar{\gamma}}(e) < \infty$ holds for $\forall~e$. 
   % Lemma~2.5 of \cite{montes1994average} shows that under condition 1-3), the optimal discounted cost $J_{\alpha}^\ast$ satisfies 
   % \begin{eqnarray*}
   %     J_{\alpha}^\ast (e) = \min_{\delta \in \AC(x)} \big\{ g(e,\delta) + \alpha \int J_{\alpha}^\ast(y)\PC({\rm d}y \vert e, \delta)\big\}, \ \forall e \in \XC
   % \end{eqnarray*}
   For arbitrary but fixed state $\varepsilon$, select  $M=\theta$, the remaining statement of condition 4) holds. 
For condition 5), let $T_0 = \inf_{t > 0}\{t: e_t = \varepsilon \}$, we have 
\begin{eqnarray*}
    J_{\alpha}^\ast (e)   &\hspace{-0.5em}\le \hspace{-0.5em}& \Eb[\sum_{k=0}^{T_0} \alpha^k g(e_k,1)\vert e_0=e] + \Eb[ \alpha^{T_0}   \vert e_0=e] J_{\alpha}^\ast (\varepsilon) \nonumber \\
   &\hspace{-0.5em}\le \hspace{-0.5em}& \Eb[\sum_{k=0}^{\infty
   }   \alpha^k  g(e_k,1)\vert e_0=e] + J_{\alpha}^\ast (\varepsilon).
\end{eqnarray*}
Thus, we select $b(e) = \Eb[\sum_{k=0}^{\infty}   \alpha^k g(e_k,1)\vert e_0=e] =  \frac{\theta}{1-\alpha}$ to satisfy $h_\alpha(e) \le b(e)$. Additionally, since $    J_{\alpha}^\ast (e) \le J_{\alpha}^{\bar{\gamma}}(e) < \infty$ establishes for all $e$, we have $J_{\alpha}^\ast(e) - J_{\alpha}^\ast(\varepsilon) \ge -N$ for some positive $N$. 
Moreover, $\int b(y)\PC({\rm d}y\vert e, \delta) < \infty$  establishes as $b(\cdot)$ is bounded. Thus, condition 5) holds.   

The definition of equicontinuity and the remaining proof for condition 6) are provided in the Appendix, where we prove that the family of functions $h_{\alpha_n}$ and $J_{\alpha_n}$ are pointwise equicontinuous. Additionally, since $h_{\alpha_n} (e)$ is bounded for $\forall~e\in\XC$, the closure of the set $h_{\alpha_n} (e)$ is compact. Then by Ascoli theorem \cite{sanchis1998note}, there is a subsequence $\{h_{n_k}\}$ of $\{h_{\alpha_n} \}$ that converges pointwisely to a continuous function $h$, and the convergence is uniform on each compact subset of $\XC$. This guarantees the existence of a measurable function $h$. 
   \hfill $\blacksquare$
\begin{remark}\label{remark:MDP}
In Bellman equation \eqref{eq:bellman}, $j^\ast= J^\ast(e) = J^\ast$ for all $e\in \XC$, is the minimum average cost, which means that the solution of the average-cost optimization problem is independent of the initial value $e$.  The error dynamics $\{e_k\}_k$ is a $\delta_k$-controlled Markov chain on $\mathbb{R}^n$, where $e_k =\varepsilon$ is the fixed recurrent state.  Then, the differential cost $h(e)$ is interpreted as the difference between the expected cost to reach fixed state $\varepsilon$ from state $e$ for the first time and the expected cost if the stage cost were $j^\ast$ rather than $g(e,\delta)$, i.e., $h(e) = J^\ast(e) - J^\ast(\varepsilon) $ \cite{bertsekas2005dynamic}. Additionally, given the symmetric nature of the Gaussian process distributions, it is natural to select the fixed state as $\varepsilon=\mathbf{0}$.
\end{remark}

Proposition~\ref{proposition:bellman} shows that the optimal solution $(J^\ast,h,\gamma^\ast)$ to the Bellman equation \eqref{eq:bellman} exists, and the optimal scheduling law $\gamma^\ast$ is stationary. 
The following sections are to solve the optimal solution triplet $(J^\ast,h,\gamma^\ast)$. 

\subsection{Compute $(J^\ast,h)$}\label{sec:value iteration}

In this section, we use a value iteration algorithm to obtain the pair $(J^\ast,h)$ and explore its truncated version to reduce the computational complexity of the value iteration process. Notably, the value iteration result is independent of the specific form of the optimal scheduling law.

Define the total expected $t$-stage cost under the stationary law $\gamma$ as $    J_t^\gamma(e):= \Eb\big[\sum_{k=0}^{t-1}g(e_k,\delta_k)|e_0 = e\big]$ 
with $\delta_k = \gamma(e_k)$, for $ k = 0,\dots,t-1$. Let $J_0^\ast(e)=0$, and $J_t^\ast(e):=\inf_{\gamma \in \Gamma}J_t^\gamma(e)$ be the optimal $t$-stage cost,  for all $e\in\XC$ and $t \ge 1$.  
% Note that if $(J^\ast, h,\gamma^\ast)$ is a  solution to average-cost optimality equation \eqref{eq:bellman}, so as $(J^\ast, h+N,\gamma^\ast)$ for any constant $N \in \mathbb{R}$. Thus, without loss of generality, let $h(e) = 0$ for all $e \in \XC$. 
We apply the Bellman operator $\mathcal{T}$ to the cost $J_t^\gamma(e)$ and obtain the dynamic equation in "forward" form:  \begin{align}\label{eq:Bellman h=0}
    J_t^\ast(e) &= \mathcal{T}J_{t-1}^\ast(e)\nonumber \\ 
    &=\min_{\delta=\AC(e)}\bigg\{ g(e,\delta) + \int_{\XC}J_{t-1}^\ast(y)\PC({\rm d}y \vert e,\delta) \bigg\}, 
\end{align}
for $ \forall~t\ge1,~e\in \XC$. 
According to Proposition~\ref{proposition:bellman}, define two sequences:
\begin{eqnarray}\label{eq:ht1}
    h_t(e):= J_t^\ast(e) - J_t^\ast(\varepsilon), \ \lambda_t := J_t^\ast(e) - J_{t-1}^\ast(e),
\end{eqnarray}
for $ e \in \XC$. Then, the following lemma shows that the value iteration sequence
$\{(\lambda_t,h_t)\}$ 
will converge to $(J^\ast,h)$.
% Algorithm~\ref{alg:differential cost} presents the value iteration procedure to obtain the sequence $\{(\lambda_t,h_t)\}$. 
% Then Lemma~\ref{lemma:h convergence} shows that 
% $\{(\lambda_t,h_t)\}$ 
% will converge to $(J^\ast,h)$.  
% \begin{minipage}{8.2cm} 
% \RestyleAlgo{ruled}
% \SetKwComment{Comment}{/* }{ */}
% \begin{algorithm}[H]
% \caption{Value iteration of $(J^\ast,h)$} \label{alg:differential cost}
% \KwData{Initialize functions $J_t^\ast, h_t: \XC \rightarrow \mathbb{R}$ with $J_0^\ast(e) = h_0 (e) = 0$, for $\forall~e \in \XC$, $\lambda_t: \XC \rightarrow \mathbb{R}$ with $\lambda_0 = 0$; \\ 
% \hspace{1cm}iteration step $t$, \ fixed state $\varepsilon = \mathbf{0}$;\\
% \hspace{1cm}a small positive scalar $ \kappa  $; }
% \KwResult{$h_{t+1} $,\ $\lambda_t$} 
% \While{$\big|h_{t+1}(e) - h_{t}(e)\big| > \kappa$}{
% \begin{algorithmic}
% \STATE \eqref{eq:Bellman h=0}, \eqref{eq:ht1}
% \end{algorithmic}}
% \end{algorithm}
% \end{minipage}
% \vspace{0.5em}
\begin{lemma}\label{lemma:h convergence}
The sequence $ \{(\lambda_t,h_t)\}$ obtained by \eqref{eq:Bellman h=0}, \eqref{eq:ht1} converges to  $(J^\ast, h ) $, which is the solution of \eqref{eq:bellman}. 
\end{lemma}
\textit{Proof}. 
For a deterministic Markov policy $\gamma  \in \Gamma$, denote $\PC^t(\cdot | e,\gamma)$   as the $t$-step transition kernel and $\PC^t(\cdot \vert e,\gamma):= \PC(\cdot \vert e,\gamma)\PC^{t-1}(\cdot \vert e,\gamma)$. 
According to Theorem 5.6.3 of \cite{hernandez2012discrete},
we need to verify:
\begin{enumerate}
    \item the sequence $\{J_t^\ast  \}$ is equicontinuous. 
    \item for any deterministic stationary policy $\gamma \in \Gamma$, there exists a probability $\PC_\gamma$ on $\XC$ such that $\PC^t(\cdot \vert e,\gamma)$ weakly converge to $\PC_\gamma$ as $t$ approaches infinity, for all $e \in \XC$ and $\PC_{\gamma}(\mathcal{G}) >0$ for every open set $\mathcal{G}$. 
    \item there is a function $L: \XC \rightarrow \mathbb{R}$ such that 
    \begin{eqnarray}\label{eq:convergence condition3}
        \int b(y) \PC^t\big({\rm d}y \vert e, (\gamma_t,\gamma_{t-1},\dots,\gamma_1)\big) \le L(e),  
    \end{eqnarray}
    for $\forall~t\ge 1$ and $e\in \XC$, 
    where $\gamma_1,\dots,\gamma_t$ is the policy sequence, $b(\cdot)$ is as in Proposition~\ref{proposition:bellman}. 
\end{enumerate}
Since the proof of condition 6) in Proposition~\ref{proposition:bellman} also applies to the case $\alpha_t =1$, condition 1) holds.  As in Remark 5.6.2 of
\cite{hernandez2012discrete}, condition~2) is equivalent to verify
\begin{eqnarray}\label{eq:contraction condition}
    \|\PC(\cdot|e,\delta) - \PC(\cdot | e',\delta')\|_{tv} \le 2 \beta 
\end{eqnarray}
with $0<\beta<1$, for $\forall~(e,\delta), (e',\delta') \in \KC$. 
This is because \eqref{eq:contraction condition} implies that  $\|\PC^t(\cdot|x,\gamma) - \PC_\gamma(\cdot)\|_{tv} \le 2 \beta^t$, where $2 \beta^t$ approaches zero when $t$ goes to infinity. Furthermore, the left-hand side of \eqref{eq:contraction condition} is written as
\begin{eqnarray}
    &&\| \PC(\cdot | e,\delta) - \PC(\cdot | e',\delta')\|_{tv} \nonumber \\
  \eq 2 \sup_{\mathcal{B}\in \mathbb{R}^n}\big|\PC( \mathcal{B}| e,\delta) -  \PC ( \mathcal{B} | e',\delta') \big| \nonumber \\ 
  \eq \int_{\XC}\big|p(y|e,\delta)-p(y|e',\delta')\big|{\rm d}y \nonumber \\ 
&\hspace{-0.5em}\leq\hspace{-0.5em}& \int_{\XC} p(y|e,\delta)+p(y|e',\delta') \nonumber \\
  &&- 2\min\{ p(y|e,\delta), p(y|e',\delta') \}{\rm d}y \nonumber \\ 
  \eq 2 - 2\bar{\beta} 
\end{eqnarray}
with $\bar{\beta} = \int_{\XC} \min\{ p(y|e,\delta), p(y|e',\delta') \}{\rm d}y$. The second equality follows from Scheffe's Theorem \cite{devroye1985nonparametric}. 
Let $\beta = 1-\bar{\beta}$. Since $0<\bar{\beta}<1$, condition 2) holds.

For condition 3), \eqref{eq:convergence condition3} holds by selecting $L(e) = b(e) = \frac{\theta}{1-\alpha}$. 
Upon verifying that conditions 1-3) are satisfied, by Theorem 5.6.3 of \cite{hernandez2012discrete}, the sequence $\{(\lambda_t,h_t)\}$ generated by iteration procedure \eqref{eq:Bellman h=0} and \eqref{eq:ht1} converges to $(J^\ast, h)$.  
\hfill $\blacksquare$

Value iteration over an unbounded general state space  $\mathcal{X}$ is computationally challenging. Next, we solve the Bellman equation \eqref{eq:bellman} on a truncated state space and investigate the relationship between the solution of this truncated problem and the original solution $(J^\ast,h)$. 
Define a bounded and convex state space  $\bar{\XC}$.   
Denote $(\bar{J}^\ast,\bar{h})$  as the optimal solution to the Bellman equation \eqref{eq:bellman} on the truncated space $\bar{\XC}$. Restrict the solution pair $(J^\ast, h)$ on $\bar{\XC}$, and denote it as $(J_{\bar{\XC}}^\ast,h_{\bar{\XC}})$. 

\begin{proposition}\label{proposition:truncated MDP}
Consider the Bellman equation \eqref{eq:bellman} on state space $e \in \bar{\XC}$, we have $  (\bar{J}^\ast,\bar{h}) =(J_{\bar{\XC}}^\ast,h_{\bar{\XC}})$. 
\end{proposition}
\textit{Proof}. 
With a slight abuse of notation, we still use $\PC$ to denote the transition kernel on $\bar{\XC} \times \AC$. 
Similar to Proposition~\ref{proposition:bellman}, 
for  the truncated problem \cite{hernandez2012discrete} with $e \in \bar{\XC}$, we have 
\begin{eqnarray}\label{eq:truncated bellman 1}
  \bar{J}^\ast+  \bar{h}(e) = \min_{\delta \in \AC(e) } \left\{g(e,\delta) + \int \bar{h}(y) \PC\big({\rm d} y \vert e,\delta\big) \right\}. 
\end{eqnarray}
Since \eqref{eq:truncated bellman 1} has the same form as the Bellman equation \eqref{eq:bellman}, 
$(\bar{J}^\ast, \bar{h},\gamma^\ast)$ is also a solution to \eqref{eq:bellman} on state space $\bar{\XC}$. Following the same argument, so as $(J_{\bar{\XC}}^\ast,h_{\bar{\XC}},\gamma^\ast)$. 

Since
the transition kernel on $\bar{\XC}$ satisfies  \eqref{eq:contraction condition}, the Bellman operator $\TC$ is a span contraction on $\bar{\XC}$. By Banach's fixed-point theorem \cite{smart1980fixed}, there exists a fixed point solution that satisfies \eqref{eq:bellman} for all admissible $e$, and this solution is unique.
Thus, we obtain  $(\bar{J}^\ast,\bar{h})= (J_{\bar{\XC}}^\ast,h_{\bar{\XC}})$.  \hfill $\blacksquare$ 
\begin{remark}
Proposition~\ref{proposition:truncated MDP} shows that the solution of the optimization problem on truncated state space is the same as the truncated solution of the original optimization problem, i.e., $(\bar{J}^\ast,\bar{h})= (J_{\bar{\XC}}^\ast,h_{\bar{\XC}})$. 
As will be shown in Theorem~\ref{theorem:stability VoI}, the scheduling policy to be designed restricts the error dynamics $e_k$ within a bounded state space. Consequently, when employing the value iteration approach, it suffices to calculate $(J_{\bar{\XC}}^\ast,h_{\bar{\XC}})$ on $\bar{\mathcal{X}}$, instead of $(J^\ast,h)$ on $\mathcal{X}$, thereby alleviating computation burden.
\end{remark}

\subsection{Optimal stationary scheduling}\label{sec:optimal scheduling}

This section will derive the optimal stationary scheduling law and analyze the stochastic stability of the error dynamics. Before presenting the main result, we provide the general definition of the VoI metric as follows.
\begin{definition}
The VoI is defined as the value that is assigned to the reduction of uncertainty from the decision maker’s perspective given a piece of information \cite{soleymani2021value}, which is expressed as $    {\rm{VoI}}_{k}:=J (\IC_{k}^{s})\vert_{\delta_{k}=0} - J(\IC_{k}^{s})\vert_{\delta_{k}=1}$, 
where
$J(\IC_{k}^{s})$ denotes the value function derived from the objective optimization problem.
\end{definition}
\begin{theorem}\label{theorem:stationary VoI}
Consider the optimization problem \eqref{eq:original opt} for system \eqref{eq:plant}. Fix the control law as the certainty equivalence controller \eqref{eq:controller}. The optimal scheduling law $\gamma^{\ast}$ for the average-cost minimization problem \eqref{eq:original opt} is given by 
\begin{eqnarray}\label{eq:VoI}
 \delta_{k} =  \gamma^{\ast}(\IC_{k}^{s}) =\left\{\begin{array}{ll}
1      & {\rm if}~ {\rm VoI}(e_k)\ge 0 \\
0      & {\rm otherwise},
 \end{array}\right.
\end{eqnarray}
where ${\rm{VoI}}(e)$ is defined as
\begin{eqnarray}\label{eq:VoI expression}
{\rm{VoI}}(e)  =  e^{\top}A^\top\Sigma Ae  + \Eb[ h (Ae  + \xi)] - \theta    - \Eb[h (\xi) ], 
\end{eqnarray} 
where the random variable $\xi $ follows the distribution $ \NC(0,\Xi)$, and $ \Eb[h(Ae+\xi )] = \int_{y\in\XC} h(y) p_{\xi}(y-Ae) {\rm d}y  $, $\Eb[h(\xi )] = \int_{y\in\XC} h(y) p_{\xi}(y) {\rm d}y$
denote the expected differential costs. 
Moreover, under the control law \eqref{eq:controller} and the scheduling law \eqref{eq:VoI}, the expected average cost of optimization objective \eqref{eq:original opt} is given as:
\begin{eqnarray}\label{eq:average cost}
       \Psi_{\infty}  \eq \tr(SW     +  A^{\top} \Sigma AP^{s}  +  \Sigma W)  + \Eb[h(\xi)].
\end{eqnarray}  
\end{theorem}
\textit{Proof}. 
We expand the right-hand side of Bellman equation \eqref{eq:bellman} and obtain
\begin{eqnarray}\label{eq:Bellman 2}
 J^{\ast} + h(e)  
&&\hspace{-2em}=  \min_{\delta \in\AC} \big\{ \delta\big( \theta      + \Eb[ h(\xi) ]\big)  \nonumber \\ 
&&\hspace{-1em}+ (1-\delta ) \big(e^{\top}A^\top \Sigma Ae +   \Eb[h(Ae+\xi)]\big)  \big\} \nonumber \\ 
&&\hspace{-2em}=  \min_{\delta \in \AC }\Big\{ (Ae)^{\top} \Sigma Ae +   \Eb\big[h(Ae+\xi)\big] \nonumber \\ 
&&\hspace{-6em}+ \delta\Big( \theta      + \Eb[ h(\xi) ] - (Ae)^{\top} \Sigma Ae -   \Eb\big[h(Ae+\xi)\big]\Big)  \Big\}.
\end{eqnarray}
To minimize the right-hand side of \eqref{eq:Bellman 2},  we choose $\delta=0$ when $\theta      + \Eb[ h(\xi) ] \ge (Ae)^{\top} \Sigma Ae +   \Eb\big[h(Ae+\xi)\big]$,   and $\delta=1$ otherwise. Thus, the VoI metric and the optimal scheduling law $\gamma^\ast$ are given by \eqref{eq:VoI expression} and \eqref{eq:VoI}, respectively.   

Substituting \eqref{eq:simplified cost 1} and \eqref{eq:error cost funcion} into \eqref{eq:Vk}, we obtain $\Psi (\gamma ) =  J(\gamma)     +   \tr\big(SW  + A^\top \Sigma A P^s+ \Sigma W\big) $.  {Furthermore, since the fixed state is selected as $\varepsilon = \mathbf{0}$, we have $h(\mathbf{0}) = 0 $ by \eqref{eq:ht1}.  Then, by setting $e = \mathbf{0}$ in \eqref{eq:Bellman 2}, we obtain the minimum average cost $J^{\ast} = \Eb\big[h(\xi)\big]$ for the optimization problem \eqref{eq:simplified opt}. 
Thus,  
the expected average cost $\Psi(\gamma^\ast)$ is given by \eqref{eq:average cost}.  
 \hfill $\blacksquare$
 
Compared to \cite{xu2004optimal}, our approach does not require the admissible error region to be bounded. Additionally, this is the first work to address the average-cost optimization problem with unbounded general state spaces and unbounded costs within the context of state-based event-triggered control.

\begin{remark}
The one-step delay assumption in networked communication is introduced to define the information structure between the controller and the scheduler therefore enabling the design separation. If we consider a corresponding delay-free case \cite{molin2013on}, then the optimization problem is formulated as a MDP with control model $(\XC,\AC,\PC,g_d)$ 
with the stage cost  $g_d(e_k ,\delta_k) = \theta \delta_k + (1-\delta_k) e_k^{\top}  \Sigma  e_k$. The state space $\XC$, the control space $\AC$, and the transition kernel remain unchanged.  This framework can also be extended to a constant $\tau$-step delay case by defining an appropriate stage cost and error dynamics expression, of which derivation see \cite{wang2021value}.
\end{remark}

In the following, we will investigate the boundedness of the error dynamics under the designed scheduling law \eqref{eq:VoI}. 
Based on it, we further analyze the stochastic stability of the error dynamics \eqref{eq:estimation mismatch}.

\begin{theorem}\label{theorem:stability VoI} 
Consider the error dynamics \eqref{eq:estimation mismatch} under the  scheduling law  \eqref{eq:VoI}, the $\delta_{k}$-controlled Markov chain $\{e_{k}\}_{k}$ is stochastically stable. 
\end{theorem} 
\textit{Proof}. 
We first show that the error dynamics under the scheduling law \eqref{eq:VoI} is bounded. 
From $J^{\ast} = \Eb[h(\xi)]$ and \eqref{eq:Bellman 2}, we have 
\begin{eqnarray}\label{eq:h bound}
    h(e) \le \theta + \Eb[h(\xi)] - J^\ast = \theta,
\end{eqnarray}
for all $e \in \XC$.  
As in condition 5) of Proposition~\ref{proposition:bellman}, $  h_{\alpha_n}(e) \ge -N$ for all $\alpha_n$ and $e\in\XC$, and the subsequence of $\{h_{\alpha_n}\}$ converges pointwisely to $h$ as $n$ goes infinity. Therefore, we have $-N \le h(e) \le \theta$, and
  $-N \le \Eb[h(e)] < \theta$ for all $e\in \XC$. 
Furthermore, it obtains $ -N \le  \Eb[h (\xi) ] + \theta -\Eb\big[ h (Ae  + \xi)\big] \le 2\theta +N$,  
for all $e \in \XC $. 
According to the scheduling law \eqref{eq:VoI}, we define an admissible region for the error dynamics, which is 
$\MC = \big\{ e_k\in \XC|(Ae_k)^\top \Sigma Ae_k  \le 2\theta +N  \big\}$. 

Since the random variable $\xi_{k}$ has a continuous probability density function,  the Markov chain $\{e_{k}\}_{k}$ is aperiodic and $\phi$-irreducible, where $\phi$ denotes the Lebesgue measure.   The stability condition is based on Foster's criterion for stochastic stability \cite{meyn2012markov,molin2014suboptimal}. Consider the Lyapunov candidate $V(e_{k}) = e_{k}^{\top}e_{k}$, define the drift operator as $\Delta V(e_{k}) = \Eb[V(e_{k+1}) \vert e_{k}] -V(e_{k})$.   Substituting  \eqref{eq:estimation mismatch} into $\Delta V(e_{k})$, we obtain 
\begin{eqnarray}\label{eq:drift operator}
 \Delta V(e_{k}) 
\le \Eb\big[  (1-\delta_{k}) \bar{\rho}(A^{\top}A)\|e_{k}\|^{2} + \|\xi_{k}\|^{2} \big] -\|e_{k}\|^{2}.
\end{eqnarray}
When it satisfies the triggering condition \eqref{eq:VoI}, we have $\delta_{k}=1$, and thus the right-hand side of \eqref{eq:drift operator} is further written as $  \tr(\Xi)-\|e_{k}\|^{2}$. 
Define a region $ \DC = \big\{  e_k  \in \XC \vert \|e_k \|^{2} \le \sqrt{\tr(\Xi) + 1} \big\}$.  
Then, the Foster's criterion $\Delta V(e_{k}) < -1$ establishes for $e_{k} \in \XC \backslash (\MC \bigcup \DC)$. Thus, we conclude that the error dynamics \eqref{eq:estimation mismatch} under the VoI-based scheduling law  \eqref{eq:VoI} is stochastically stable. \hfill $\blacksquare$

\subsection{Threshold and quadratic structure }\label{sec:threshold structure}
This section analyzes the structure of the VoI metric function \eqref{eq:VoI expression}, focusing on its symmetry and monotonicity properties. Building on these findings, we demonstrate that the resulting scheduling law \eqref{eq:VoI} exhibits a threshold structure and takes a quadratic form.

The following proposition presents the symmetry of the VoI function \eqref{eq:VoI expression} and the scheduling law \eqref{eq:VoI}.
\begin{proposition}\label{proposition:he symmetry}
Let $J_0^\ast(e)= 0$ for all $e\in \XC$. We have that  $J_t^\ast(e)$, $h(e)$, $\Eb[h(Ae+\xi)]$, and the optimal scheduling law \eqref{eq:VoI} are symmetric in $e \in \XC$. 
\end{proposition}
\textit{Proof.}   
We first prove that the Bellman operator $\TC$ preserves the symmetry of the function in \eqref{eq:Bellman h=0}.  Given a symmetric function $J_t^\ast(e)$, we have $\int_{\XC} J_t^\ast(y)\PC({\rm d}y \vert e,\delta) = \int_{\XC} \{  \delta J_t^\ast(y) p_{\xi}(y)  + (1-\delta )J_t^\ast (y) p_{\xi}(y-Ae ) \} {\rm d}y$. 
Since the probability density function 
$p_\xi(\cdot)$ is symmetric, we have that $\int_{\XC}   J_t^\ast(y) p_{\xi}(y){\rm d}y  $ is symmetric. Additionally, let $z:= -y$, we have 
\begin{eqnarray*}
 &&   \int_{\XC} J_t^\ast (y) p_{\xi}(y-Ae ) {\rm d}y  =  \int_{\XC} J_t^\ast (-z) p_{\xi}(-z-Ae ) {\rm d}z \nonumber \\
    \eq \int_{\XC} J_t^\ast (z) p_{\xi}(z+Ae ) {\rm d}z  = \int_{\XC} J_t^\ast (y) p_{\xi}(y+Ae ) {\rm d}y, 
\end{eqnarray*}
where the first equality holds as $\XC$ is symmetric, the second equality holds as $p_\xi(\cdot)$ and $J_t^\ast(\cdot)$ are symmetric. Thus,  we conclude that $\int_{\XC}J_t^\ast(y)\PC({\rm d}y \vert e,\delta)$ is symmetric in $e \in 
\XC$ no matter $\delta =0$ or $1$.  Furthermore, since $g(e,\delta)$ is also a symmetric function, we obtain that the Bellman operator $\TC$ preserves the symmetry during the iteration process. Moreover, the initial value of the iteration is $J_0^\ast(e) = 0$ for all $e \in \XC$, which is also symmetric. Thus, $J_t^\ast(e)$ is symmetric in $e \in \XC$ for all $t$. 

As in \eqref{eq:ht1} and Lemma~\ref{lemma:h convergence}, $h_t(e) = J_t^\ast(e) - J_t^\ast(\mathbf{0})$ and will converge to $h(e)$. Thus,  $h_t(e)$, for all $t$, and its convergence $ h(e)$ are symmetric in $e\in \XC$. 
The next is to show that  $\Eb\big[h(Ae+\xi)\big]$ is a symmetric function in $e \in \XC$. Let $z:=-\xi$, we have 
\begin{eqnarray}
  &&  \Eb\big[h(Ae+\xi)\big] = \Eb\big[h(-Ae-\xi)\big] \nonumber \\
    \eq \int_{\XC} h(-Ae-\xi)p_{\xi}(\xi){\rm d}\xi = \int_{\XC} h(-Ae+z)p_{\xi}(z){\rm d}z\nonumber \\
    \eq \Eb\big[h(-Ae+z)\big] = \Eb\big[h(-Ae
    +\xi)\big],
\end{eqnarray}
where the first equality is from $h$ is symmetric, the third equality is from  $\XC$ and $p_\xi(\cdot)$ are symmetric. 
Moreover, the remaining terms of the VoI function \eqref{eq:VoI expression}, i.e., $\Eb[h(\xi)]$ and $(Ae)^\top\Sigma Ae$,  are symmetric functions in $e$. Thus, we obtain that the VoI function \eqref{eq:VoI expression} and the VoI-based optimal scheduling law \eqref{eq:VoI} are symmetric. 
\hfill $\blacksquare$

Analyzing the monotonicity of the VoI function with a general matrix $A$ is challenging, as seen in the iteration process \eqref{eq:Bellman h=0}. To tackle this issue, we introduce a diagonalizability assumption and reformulate the MDP model in a transformed state space. We then examine the monotonicity property of the VoI function in the transformed state space. 
  
\begin{assumption}\label{ass:diagonal}
    Assume that the system matrix $A$ is diagonalizable. 
\end{assumption}

By Assumption~\ref{ass:diagonal}, there exist an invertible matrix $U$ and a diagonal matrix $\Lambda$ such that $U^{-1}AU = \Lambda$.  Denote $s_k: = U^{-1}e_k \in \XC$ for all $k$. 
The error dynamics \eqref{eq:estimation mismatch} is transformed to $$    s_{k+1} =(1-\delta_k) \Lambda s_k + \zeta_k$$ 
with $\zeta_k=U^{-1}\xi_k$.
The optimization problem \eqref{eq:simplified opt} is reformulated as a MDP model $(\XC,\AC,\PC_s,g_s)$ with the stage cost $$g_s(s_{k} ,\delta_{k}) = \theta \delta_{k} + (1-\delta_{k}) s_{k}^{\top}\Lambda U^\top \Sigma U \Lambda s_k,$$  
and the stochastic kernel $\PC_s$ on $(s,\delta) \in \XC \times \AC$: 
\begin{eqnarray}\label{eq:transition s}
\PC_s({\rm d} y \vert s , \delta ):\hspace{-0.5em}\eq p_s(y\vert s,\delta){\rm d}y \nonumber \\
\eq \big[ \delta p_{\zeta}(y)   + (1-\delta ) p_{\zeta}(y-\Lambda s )\big] {\rm d}y,  
\end{eqnarray}
where $p_s$ is the transition probability function from the current state $(s,\delta):= (s_k,\delta_k)$ to the next state $y: = s_{k+1}$, and
$p_{\zeta}$ is the probability density function of the random variable $\zeta \sim \NC\big(0, U^{-1}\Xi (U^{-1})^\top\big)$. Previous results on MDP model $(\XC,\AC,\PC,g)$ hold for arbitrary $A$, and thus also hold for MDP model $(\XC,\AC,\PC_s,g_s)$. With a slight abuse of notations, we still use $h$, $J_t^\ast$ to denote the relative cost and value iteration sequence on $s$. Accordingly, the Bellman optimality equation on $s \in \XC$ is written as 
\begin{eqnarray}\label{eq:Bellman s}
 J^{\ast} + h(s)  
&&\hspace{-2em}=  \min_{\delta \in\AC} \big\{ \delta\big( \theta      + \Eb\big[ h(\zeta) \big]\big)  \nonumber \\ 
&&\hspace{-2.5em}+ (1-\delta ) \big(s^{\top}\Lambda U^\top \Sigma U \Lambda s +   \Eb\big[h(\Lambda s+\zeta)\big]\big)  \big\} 
\end{eqnarray}
with $ \Eb\big[h(\Lambda s+\zeta )\big] = \int_{y\in\XC} h(y) p_{\zeta}(y-\Lambda s) {\rm d}y$ and 
$  \Eb[h(\zeta )] = \int_{y\in\XC} h(y) p_{\zeta}(y) {\rm d}y$. 
The optimal scheduling law  depending on $s$ 
 is given as 
\begin{eqnarray}\label{eq:VoI s}
\delta_{k} =  \gamma^{\ast}(\IC_{k}^{s}) =\left\{\begin{array}{ll}
1      & {\rm if}~ {\rm VoI}(s_k)\ge 0 \\
0      & {\rm otherwise},
\end{array}\right.
\end{eqnarray}
where the VoI function is given by:
\begin{eqnarray}\label{eq:VoI expression s}
{\rm{VoI}}(s)  = s^{\top}\Lambda U^\top \Sigma U \Lambda s  + \Eb\big[ h (\Lambda s  + \zeta)\big]  - \theta   - \Eb[h (\zeta) ]. 
\end{eqnarray}

In the following, we will analyze the monotonicity of the differential cost $h(s)$ and the expectation $\Eb[h(\Lambda s +\zeta)]$.
For notational simplicity, denote  $\NC := \{1,\dots,n\}$. Denote $s_i \in \mathbb{R}$ as the $i$-th componnent of the vector $s$, for $i \in \NC$. 
\begin{lemma}
\label{lemma:monotonicity Eh}
Let Assumption~\ref{ass:diagonal} hold. For $i \in \NC$, if  $h(s)$ is monotonically nonincreasing in $s_i$, for $s_i < 0$, and monotonically nondecreasing in $s_i$, for $s_i \ge  0$, then so as $\Eb[h(\Lambda s+\zeta)]$. 
\end{lemma}

The proof of Lemma~\ref{lemma:monotonicity Eh} is provided in the Appendix. 
\begin{proposition}
\label{proposition:H J mono}
Let $J_0^\ast(s)= 0$ for all $s\in \XC$. Let Assumption~\ref{ass:diagonal} hold. For $i \in \NC$, we have that $J_t^\ast(s) $, $h(s)$ and $\Eb[h(\Lambda s+\zeta)]$ are monotonically nonincreasing in $s_i$, for $s_i \le 0 $; and are monotonically nondecreasing in $s_i$, for $s_i \ge 0 $. 
\end{proposition} 
\vspace{-1em}\noindent
\textit{Proof}. 
From Proposition \ref{proposition:he symmetry}, $J_t^\ast(s)$, $h(s)$, $\Eb\big[h(\Lambda s +\zeta)\big]$ and ${\rm{VoI}}(s)$ are even in $s \in \XC$.
Thus, we first focus on the case $s_i \ge 0$, for $i\in \NC$. 
Define a vector $\vec{s}_{i}  = \big[\tilde{s}_1,\dots, s_i, \dots,\tilde{s}_n \big]^\top \in \XC$ with $\tilde{s}_j$ being some fixed scalars, for $j \in \NC \backslash i$. Similarly,  define a vector $\vec{\epsilon}_{i} = \big[\tilde{s}_1,\dots, \epsilon_{i}, \dots,\tilde{s}_n \big]^\top \in \XC$, where $\epsilon_{i} > s_{i}$.
Then,  we use the mathematical induction method to prove that  $J_t^\ast(\vec{\epsilon}_i) \ge J_t^\ast(\vec{s}_i)$ holds for $\epsilon_{i} > s_{i} \ge 0$  and all $t \ge 0$. 

For $t \ge 0$, suppose that $J_t^\ast(\vec{\epsilon}_{i}) \ge J_t^\ast(\vec{s}_{i})$, the next is to verify $J_{t+1}^\ast(\vec{\epsilon}_{i}) \ge J_{t+1}^\ast(\vec{s}_{i})$. Firstly, given that $(A,B)$ is stablizable and $(A,Q^{\frac{1}{2}})$ is detectable, the matrix $S$ solved from algebraic Riccati equation \eqref{eq:Riccati matrix} is positive semidefinite \cite{hager1976convergence}. Moreover, since the matrix $R$ is positive definite, matrices $\Sigma$ and $\Lambda U^\top  \Sigma U \Lambda$ are also 
positive definite by their definition. Then, for $ \epsilon_i > s_i \ge0$,  we have
\begin{eqnarray}\label{eq:delta=0, mono}
&& \vec{\epsilon}_{i}^{\top} \Lambda U^\top  \Sigma U \Lambda \vec{\epsilon}_{i} + \Eb\big[J_t^\ast(\Lambda \vec{\epsilon}_{i}+\zeta )\big]  \nonumber \\
&& \hspace{-1em}\ge \vec{s}_{i}^{\top}\Lambda U^\top  \Sigma U \Lambda \vec{s}_{i} + \Eb\big[J_t^\ast(\Lambda \vec{s}_{i}+\zeta )\big], 
\end{eqnarray}
where $\vec{\epsilon}_{i}^{\top} \Lambda U^\top  \Sigma U \Lambda \vec{\epsilon}_{i}  \ge \vec{s}_{i}^{\top}\Lambda U^\top  \Sigma U \Lambda \vec{s}_{i}$ as $ \Lambda U^\top  \Sigma U \Lambda$ is positive definite. The remaining of \eqref{eq:delta=0, mono} holds by Lemma~\ref{lemma:monotonicity Eh}.
Define $\Phi_{t}(\varrho) = \min\big\{ \theta + \Eb[J_t^\ast(\zeta)], \varrho\big\}$ with $\varrho$ being a properly chosen scalar. 
Since $\Phi_{t}(\varrho)$ is monotonically nondecreasing in $\varrho$, it obtains 
\begin{eqnarray*}
J_{t+1}^\ast(\vec{\epsilon}_{i}) 
&\hspace{-0.8em}= \hspace{-0.8em}& \Phi_t\big(\vec{\epsilon}_{i}^{\top} \Lambda U^\top  \Sigma U \Lambda\vec{\epsilon}_{i} + \Eb[J_t^\ast(\Lambda \vec{\epsilon}_{i}+\zeta )] \big)  \\
&\hspace{-0.8em} \ge \hspace{-0.8em}& \Phi_t\big(\vec{s}_{i}^{\top} \Lambda U^\top  \Sigma U \Lambda \vec{s}_{i} + \Eb[J_t^\ast(\Lambda \vec{s}_{i}+\zeta )]   \big) \hspace{-0.2em}= \hspace{-0.2em} J_{t+1}^\ast(\vec{s}_{i}).
\end{eqnarray*}
When $\vec{\epsilon}_{i}$ satisfying the triggering condition \eqref{eq:VoI s},  we have 
\begin{eqnarray*}
J_{t+1}^\ast(\vec{\epsilon}_{i}) 
&\hspace{-0.8em} = \hspace{-0.8em}& 
\theta + \Eb[J_t^\ast(\zeta)] \\
&\hspace{-0.8em} \ge \hspace{-0.8em}&\Phi_t\big(\vec{s}_{i}^{\top} \Lambda U^\top  \Sigma U \Lambda\vec{s}_{i} + \Eb\big[J_t^\ast(\Lambda \vec{s}_{i}+\zeta)\big]\big)
 \hspace{-0.2em} = \hspace{-0.2em} J_{t+1}^\ast(\vec{s}_{i}).
\end{eqnarray*}
Here is the induction proof. 
Additionally,  since $J_0^\ast(s) = 0$ for all $s$, we have 
\begin{align*}
J_{1}^\ast(\vec{\epsilon}_{i}) 
&=\Phi_1\big(\vec{\epsilon}_{i}^{\top} \Lambda U^\top  \Sigma U \Lambda\vec{\epsilon}_{i}    \big) \\
&\ge \Phi_1\big(\vec{s}_{i}^{\top} \Lambda U^\top  \Sigma U \Lambda\vec{s}_{i}     \big) = J_{1}^\ast(\vec{s}_{i}), 
\end{align*}
for $\epsilon_i > s_i \ge 0$.   
When $\vec{\epsilon}_{i}$ satisfying \eqref{eq:VoI s},  we have $J_{1}^\ast(\vec{\epsilon}_{i}) =  \theta   \ge \Phi_1(\vec{s}_{i}^{\top} \Lambda U^\top  \Sigma U \Lambda \vec{s}_{i} ) =J_{1}^\ast(\vec{s}_{i})$.
Thus, we have that  $J_{t}^\ast(\vec{\epsilon}_{i}) \ge J_{t}^\ast(\vec{s}_{i})$ for $\epsilon_i > s_i \ge 0$, $i \in \NC$ and $ t \ge 0$.  Following similar arguments, one can verify  $J_{t+1}^\ast(\vec{\epsilon}_{i}) \ge J_{t+1}^\ast(\vec{s}_{i})$ for $\epsilon_i < s_i < 0$. This induction proof demonstrates that $J_{t+1}^\ast(\vec{\epsilon}_{i})$ always increases faster than $J_{t+1}^\ast(\vec{s}_{i})$ during the iteration process. Thus, we obtain that $J_t^\ast(s) $ is monotonically nonincreasing in $s_i$, for $s_i \le 0 $; and is monotonically nondecreasing in $s_i$, for $s_i \ge 0 $.

Since $h_t(s) = J_t^\ast(s)-J_t^\ast(\mathbf{0})$, for all $t$ and $s \in \XC$, and $h(\cdot)$ is its convergent result, we obtain the monotonicity property of $h(s)$ stated in Proposition~\ref{proposition:H J mono}.  
The monotonicity property of $\mathbf{E}(\Lambda s+ \zeta)$ follows from Lemma~\ref{lemma:monotonicity Eh}. 
\hfill $\blacksquare$
\begin{remark}\label{remark:H J mono}
Proposition~\ref{proposition:H J mono} shows that  value functions $J_t^\ast(s)$, $h(s)$ and $\Eb[h(\Lambda s+\zeta)]$  nondecrease with an increasing $|s_i|$ in each dimension, and their minima are attained at  $s=\mathbf{0}$. This aligns with intuition, as we consider a quadratic cost function as in \eqref{eq:Vk}, and $h(s)$ represents the optimal cost of an associated stochastic shortest path problem starting from the nontermination states $s \in \XC $, see \cite{bertsekas2005dynamic}. Additionally,  $h(s)$ and $\Eb\big[h(\Lambda s+\zeta)\big]$ nondecrease with the increasing $|s_i|$ in each dimension, implying that they are \textit{quasi-convex} functions.
\end{remark}
\begin{remark}
Lemma~\ref{lemma:monotonicity Eh} and Proposition \ref{proposition:H J mono} hold for any random variable $\xi_k$, provided it has a symmetric and unimodal probability density function. 
Additionally, Assumption~\ref{ass:diagonal} serves as a sufficient condition for proving Lemma~\ref{lemma:monotonicity Eh}. 
The system parameter $A$ being a diagonal matrix or a scalar are special cases of a diagonalizable $A$. 
The monotonicity result in \cite{lipsa2011remote} is a special case of our findings, specifically for a first-order system where $A$ is a scalar.
\end{remark}

Building on the symmetry and monotonicity properties of the VoI function, the following theorem analyzes the structure of the scheduling law \eqref{eq:VoI}.
 \begin{theorem}\label{theorem:threshold type}
Let Assumption~\ref{ass:diagonal} hold. The optimal scheduling law \eqref{eq:VoI} exhibits a threshold structure and takes a quadratic form:
\begin{eqnarray}\label{eq:threshold triggering}
    \hat{\gamma}^\ast(\IC_k^s) = \left\{\begin{array}{ll}
    1     &  {\rm if}~ e_k^\top A^\top \Sigma A e_k \ge \eta^\ast \\
     0    &   {\rm otherwise}, 
    \end{array} \right.
\end{eqnarray}
where $ \eta^\ast = \theta + \Eb[h(\zeta)]-\Eb\big[h(\Lambda s^\ast + \zeta)\big] \le \theta  $ is a positive scalar, with $s^\ast  \in \XC$ such that ${\rm VoI}(s^\ast) = 0 $.   
\end{theorem}
\textit{Proof.} 
According to the Bellman equation \eqref{eq:Bellman s}, define a function $L(s,\delta): \XC \times \AC \rightarrow \mathbb{R} $:  
\begin{eqnarray}\label{eq:submodular function}
    L(s,\delta):\hspace{-0.5em}\eq  (1-\delta ) \big(   s^\top  \Lambda U^\top \Sigma U \Lambda s +   \Eb[h(\Lambda s+\zeta)]\big) \nonumber \\
    &&+ \delta( \theta      + \Eb[ h(\zeta) ]). 
\end{eqnarray}
Note that $L(s,\delta)$ is symmetric in $s \in \XC$. We first prove that 
\begin{eqnarray}\label{eq:submodular inequality}
    L(\vec{\epsilon}_i,\delta_1) + L( \vec{s}_i,\delta_2) \le L(\vec{\epsilon}_i,\delta_2) + L(\vec{s}_i,\delta_1) 
    % L(\vec{\epsilon}_i,1) + L( \vec{e}_i,0) \le L(\vec{\epsilon}_i,0) + L(\vec{e}_i,1)
 \end{eqnarray}
holds 
for $\epsilon_i > s_i \ge 0 $ and $\delta_1 =1$, $\delta_2=0$. 
Substituting \eqref{eq:submodular function} into \eqref{eq:submodular inequality} and eliminating identical terms on both sides, it remains to verify 
\begin{eqnarray}\label{eq:submodular inequality 1}
    &&   (1-\delta_1 ) \big(  \vec{\epsilon}_i^\top \Lambda  U^\top  \Sigma U \Lambda \vec{\epsilon}_i +   \Eb\big[h(\Lambda \vec{\epsilon}_i+\zeta)\big]\big) \nonumber \\
    &&    + (1-\delta_2 ) \big(     \vec{s}_i^\top \Lambda  U^\top \Sigma U \Lambda \vec{s}_i +   \Eb\big[h(\Lambda  \vec{s}_i+\zeta)\big]\big) \nonumber \\
      \le \hspace{-2em}&&   (1-\delta_2 ) \big(    \vec{\epsilon}_i^\top \Lambda  U^\top \Sigma U \Lambda\vec{\epsilon}_i +   \Eb\big[h(\Lambda\vec{\epsilon}_i+\zeta)\big]\big) \nonumber \\
    &&  + (1-\delta_1 ) \big(   \vec{s}_i^\top \Lambda  U^\top \Sigma U \Lambda\vec{s}_i +  \Eb\big[h(\Lambda\vec{s}_i+\zeta)\big]\big).
\end{eqnarray}
Substituting $\delta_1=1$, $\delta_2 = 0$ into \eqref{eq:submodular inequality 1}, then \eqref{eq:submodular inequality 1} holds by \eqref{eq:delta=0, mono}. Moreover, \eqref{eq:submodular inequality} can be further written as  $ L(\vec{\epsilon}_i,1) - L(\vec{\epsilon}_i,0) \le  L(\vec{s}_i,1) - L(\vec{s}_i,0)$, 
for $\epsilon_i > s_i \ge 0$.
It implies that if $L(\vec{s}_i,1) \le L(\vec{s}_i,0)$, then  $L(\vec{\epsilon}_i,1) \le L(\vec{\epsilon}_i,0)$ must holds. Namely, there exist a constant $s_i^\ast$ and the corresponding vector $\vec{s}_i^\ast = \big[\tilde{s}_1 \cdots s_i^\ast \cdots\tilde{s}_n\big]^\top$ such that,  if  $L(\vec{s}_i,1) \le L(\vec{s}_i,0)$ holds for $ s_i^\ast >0$, then $L(\vec{\epsilon}_i,1) \le L(\vec{\epsilon}_i,0)$. Following the same argument,  we  prove that \eqref{eq:submodular inequality} also holds for $\epsilon_i < s_i < 0$, and $\delta_1=1, \delta_2=0$. 
Note that $\vec{s}_i \in \XC$, for $i\in\NC$, can be arbitrary vector, and all these threshold points satisfy $L(\vec{s}_i^\ast,1) = L(\vec{s}_i^\ast,0)$, i.e., ${\rm VoI}(\vec{s}_i^\ast) = 0$. Hence, we establish that the optimal stationary scheduling law is of a threshold type.

The next is to show that the scheduling law takes a quadratic form. For clarity, we define  $\iota:=\Lambda s$. According to the VoI metric \eqref{eq:VoI expression s}, define a function $F(\iota) = \iota^\top U^\top \Sigma U \iota + \Eb[h(\iota+\zeta)] - \Eb[h(
\zeta)]- \theta$. In Proposition~\ref{proposition:H J mono}, we set $\Lambda =I$ and obtain that $\iota^\top U^\top \Sigma U \iota$, $ \Eb[h(\iota+\zeta)]$ and $F(\iota)$ are monotonically nondecreasing in $\iota_i$, for $\iota_i \ge 0 $,  and monotonically nonincreasing  in $\iota_i$ for $\iota_i < 0 $,  $i\in \NC$.             Intuitively, the region defined by $F(\iota) \le 0$ is equivalent to the region defined by $\iota^\top U^\top \Sigma U  \iota \le \eta^\ast$ with $\eta^\ast= \theta + \Eb[h(\zeta)]-\Eb[h(\iota^\ast+\zeta)]$ and $F(\iota^\ast) = 0$. Moreover, 
$\iota^\top U^\top \Sigma U  \iota =s^\top  \Lambda U^\top \Sigma U \Lambda s = e^\top A^\top \Sigma A e$. Thus, we obtain 
\eqref{eq:threshold triggering}. 

The next is to analyze the region of $\eta^\ast$. By Proposition~\ref{proposition:H J mono} and Remark~\ref{remark:H J mono}, we have  $h(s) \ge h(\mathbf{0})=0$ for all $s \in \XC$.
From \eqref{eq:h bound}, we have $0 \le h(s) \le \theta$ and $0 \le \Eb[h(s)] < \theta$ for all $s \in \XC$.  Furthermore, by Proposition~\ref{proposition:H J mono}, we have $ \Eb[h(\zeta)] \le \Eb\big[h(\Lambda s+\zeta)\big] $,  for all $s\in \XC$.
Therefore, it obtains $  0 \le  \Eb[h (\zeta) ] + \theta -\Eb\big[ h (\Lambda s  + \zeta)\big] \le \theta$, 
for all $s \in \XC $. 
Thus, from the expression of the VoI function \eqref{eq:VoI expression}, we obtain $\eta^\ast \le \theta$. 
\hfill $\blacksquare$
\begin{remark}
The threshold value $\eta^\ast$ can be computed using $h$ generated by the iteration procedure \eqref{eq:Bellman h=0}, \eqref{eq:ht1}. 
However, it is not practical to obtain an accurate iteration result on a continuous state space (for a detailed explanation, refer to Section~\ref{sec:simulation}). The threshold-based and quadratic scheduling law \eqref{eq:threshold triggering} allows us to find the optimal threshold using a brute-force search method, where the search region for the optimal threshold is provided in Theorem~\ref{theorem:threshold type}.
\end{remark}

\section{Simulation}\label{sec:simulation}
In this section, we use numerical simulation to illustrate the theoretical result. Since the difference caused by the linear transformation is trivial in theory, we chose the system matrices of the plant \eqref{eq:plant} as $A = \mathrm{diag}\{1.3,-1.1 \}$, $B = [0.1~0.1]^\top$ and $C = \mathrm{diag}\{1,1\}$. 
The Gaussian noise covariances are $W =V = 0.001\ast {\rm diag}\{1, 1\}$. The initial value is $x_{0} = [0~0]^\top$. The weighting coefficients in LQG cost function \eqref{eq:original opt} are chosen as $Q=R={\rm diag}\{1, 1\}$. 
The optimal controller gain \eqref{eq:control gain} is given as $K={\rm diag}\{ 5.4154, 2.2606
\}$.

Fig.~\ref{fig:h} shows the differential cost $h(e)$ and its expectation $\Eb\big[h(Ae+\xi)\big]$, obtained from the value iteration process \eqref{eq:Bellman h=0}, \eqref{eq:ht1}.   
The transmission price is chosen as $\theta = 0.2$. By Proposition~\ref{proposition:truncated MDP}, we choose the truncated region as $\bar{\XC} = \big\{e\in \mathbb{R}^2 \big| |e_1| \le 0.2, |e_2|\le 0.2\big\}$. As the state space is continuous, we discretize $\bar{\XC}$ into $60\ast 60$ pieces and use the round function to map the error into the corresponding region.  The contour lines of the functions $h(e)$ and $\Eb\big[h(Ae+\xi)\big]$ are shown on the right side of  Fig.~\ref{fig:h}. The contour lines do not intersect, confirming that the function is monotonically nondecreasing with $|e_i|$ for $i \in \mathcal{N}$. Fig.~\ref{fig:h} shows that $h(e)$ and $\Eb\big[h(Ae+\xi)\big]$ are both symmetric in $e$ and increase with the increasing $|e_i|$ in each dimension. This aligns with the findings of Propositions~\ref{proposition:he symmetry} and \ref{proposition:H J mono}. Moreover, Fig.~\ref{fig:h} shows that $0 \le h(e) \le \theta$ and $0 \le \Eb\big[h(Ae+\xi)\big] \le \theta$, which aligns with Theorem~\ref{theorem:threshold type}. Notably, the rounding operation inevitably introduces quantization errors. 
This quantization error also exists when calculating the expected differential cost $\Eb\big[h(Ae+\xi)\big]$, and becomes more noticeable when the covariance of random noise $\xi$ is large relative to $\theta$. These all impair the accuracy of iteration results and the corresponding scheduling performance. 
Thus, it is necessary to carefully choose the volume of the admissible region and the cardinality when discretizing the state space. As shown in Theorem~\ref{theorem:threshold type}, the VoI-based scheduling law \eqref{eq:VoI} is equivalent to the quadratic scheduling law \eqref{eq:threshold triggering}, when $A$ is diagonalizable. Thus, we demonstrate the performance of the optimal scheduling law based on \eqref{eq:threshold triggering}.
\begin{figure}[htbp]
    \centering
    \includegraphics[width=0.45\textwidth]{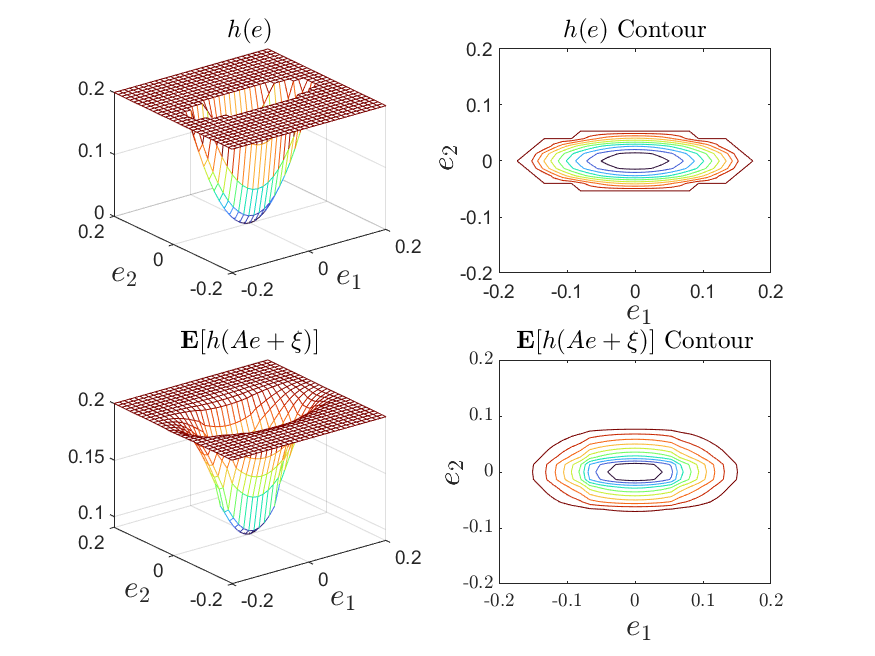}
    \caption{The differential cost $h(e)$, the expected differential cost $\Eb[h(Ae+\xi)]$ and their contour lines. } \label{fig:h}
\end{figure}

Fig.~\ref{fig:threshold} depicts the empirical sum of regulation and communication costs with different transmission prices $\theta$ under four scheduling policies.
We use $J(\gamma)$ defined in \eqref{eq:Vk} to evaluate the sum of the regulation and communication costs under the scheduling law $\gamma$. The remaining three scheduling laws used for comparison are: the greedy law   
\begin{eqnarray}\label{eq:greedy}
\delta_{k} = \gamma_{\rm G}(e_k) =  \left\{\begin{array}{ll}
1      & {\rm if}~ (Ae_{k})^{\top}\Sigma Ae_{k}  \ge  \theta \\
0      & {\rm otherwise},
 \end{array}\right.
\end{eqnarray} 
and two state-based scheduling policies: 
\begin{eqnarray}\label{eq:triggering 1}
\delta_{k} = \gamma_{\rm 1}(e_k)  = \left\{\begin{array}{ll}
1      & {\rm if}~ \|Ae_k\| \ge \eta_1^\ast \\
0      & {\rm otherwise},
 \end{array}\right.
\end{eqnarray} 
and 
\begin{eqnarray}\label{eq:triggering 2}
\delta_{k} =\gamma_{\rm 2}(e_k) =  \left\{\begin{array}{ll}
1      & {\rm if}~ \|e_k\| \ge \eta_2^\ast \\
0      & {\rm otherwise}.
 \end{array}\right.
\end{eqnarray} 
% We search the optimal thresholds using the exhaustive searching method. 
The thresholds $\eta^\ast$, defined in \eqref{eq:threshold triggering}, along with $\eta_1^\ast $ and $\eta_2^\ast$, are determined by discretizing the admissible threshold region and selecting the optimal points corresponding to the minima of the costs $J( \hat{\gamma}^\ast)$, $J(\gamma_1) $  and $J(\gamma_2) $, respectively. 
We choose the transmission price $\theta \in [0.1,5]$. The time horizon is chosen as $T=1000$. Monte Carlo simulation runs 2000 trials.   Fig.~\ref{fig:threshold} depicts the optimal thresholds  $\eta^\ast$, $\eta_1^\ast $ and $\eta_2^\ast$. It can be observed that $\eta^\ast \le \theta$, which aligns with Theorem~\ref{theorem:threshold type}. Moreover, $\eta_1^\ast, \eta_2^\ast < \eta$ as ${\rm eig}(\Sigma) >1 $. 
Fig.~\ref{fig:comparison} depicts the empirical sum of communication and regulation costs under scheduling policies \eqref{eq:threshold triggering}, \eqref{eq:triggering 1} and \eqref{eq:triggering 2}, i.e., $J(\hat{\gamma}^\ast)$, $J(\gamma_1)$ and $J(\gamma_2)$, with different multipliers. Shaded areas represent $\pm$ one standard deviation over 2000 Monte Carlo trials.  
It can be observed that the scheduling law \eqref{eq:threshold triggering} outperforms the remaining state-based scheduling policies. 

Fig.~\ref{fig:tradeoff} depicts the tradeoff between the communication rate and regulation cost under different scheduling policies. The regulation cost is measured by $\frac{1}{T } \sum_{k=0}^{T}(Ae_{k})^{\top}\Sigma Ae_{k}  $ and the communication rate is measured by $\frac{1}{T } \sum_{k=0}^{T}  \delta_{k} $. We do not include the greedy law as it has the same formulation as  \eqref{eq:threshold triggering} and thus achieves the same performance in this tradeoff. 
% By varying multiplier $\theta$, we obtain corresponding rate-regulation pairs and then line them up. 
Fig.~\ref{fig:tradeoff} shows that the state-based scheduling policies achieve better performance than the periodical one, and the optimal scheduling law in \eqref{eq:threshold triggering} yields the best performance among all the scheduling policies.  
\begin{figure}[htbp]
    \centering    \includegraphics[width=0.45\textwidth]{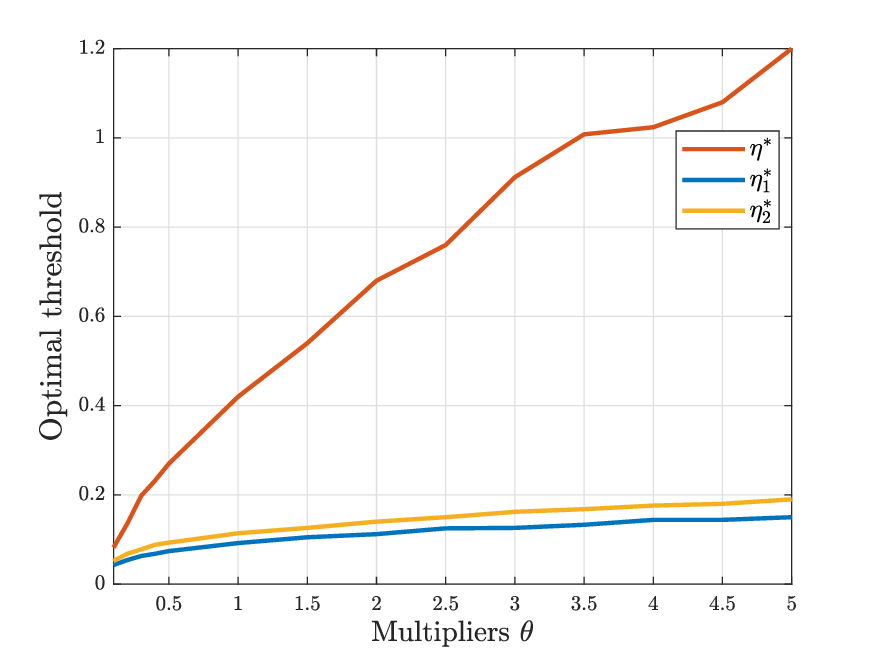}
    \caption{The optimal thresholds of optimal scheduling law $\hat{\gamma}^\ast$, and two state-based scheduling law $\gamma_1$ and $\gamma_2$ under different multiplier $\theta$.} \label{fig:threshold}
\end{figure} 
\begin{figure}[htbp]
    \centering
    \includegraphics[width=0.45\textwidth]{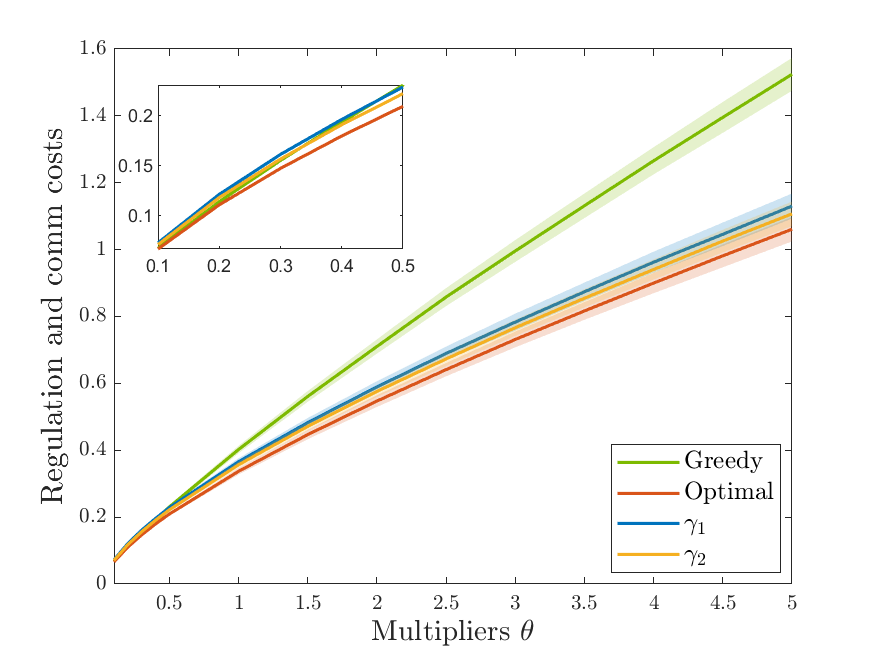}
    \caption{The regulation and communication costs \eqref{eq:Vk} achieved by the optimal scheduling \eqref{eq:threshold triggering}, the greedy scheduling \eqref{eq:greedy} and two state-based scheduling \eqref{eq:triggering 1}, \eqref{eq:triggering 2} under different multiplier $\theta$.  } \label{fig:comparison}
\end{figure}  

\begin{figure}[htbp]
    \centering
    \includegraphics[width=0.45\textwidth]{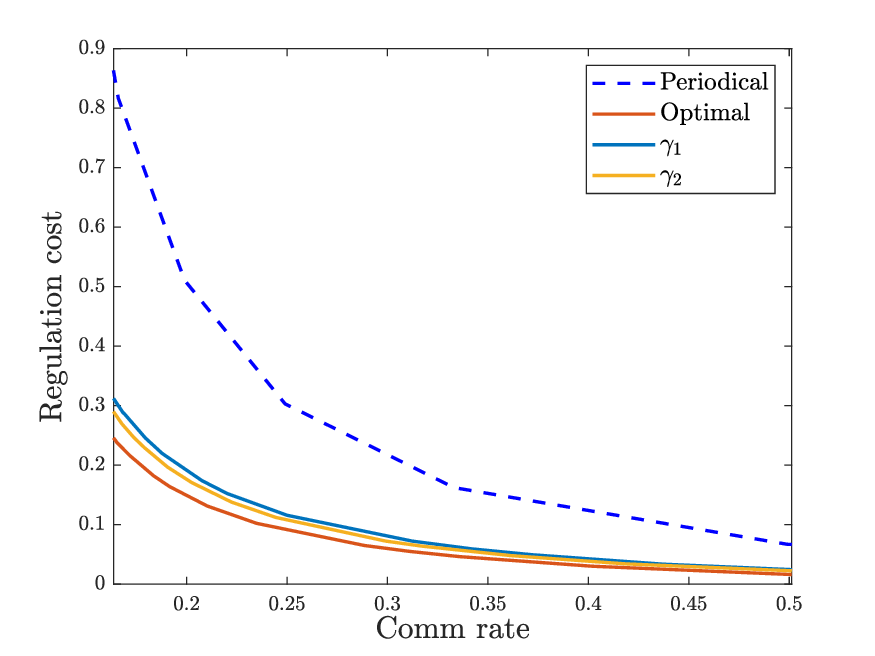}
    \caption{Tradeoff between regulation cost and communication rate under the optimal scheduling \eqref{eq:threshold triggering}, the greedy scheduling \eqref{eq:greedy} and two state-based scheduling \eqref{eq:triggering 1}, \eqref{eq:triggering 2} . } \label{fig:tradeoff}
\end{figure}   

\section{Conclusion}\label{sec:conclusion}
In this article, we designed the optimal scheduling law to minimize long-term regulation and communication costs. By fixing the control law as the certainty equivalence controller, the state-based scheduling law can be designed independently. We formulated the optimization problem as an average-cost MDP and showed that its optimal solution yields a triplet comprising an optimal stationary scheduling law based on the VoI metric, the optimal average cost, and the differential cost. These costs were computed using a value iteration approach. To reduce computational complexity, we showed that the solution could be computed on a truncated state space without compromising optimality. Additionally, the error dynamics under the designed scheduling law were shown to be bounded, ensuring the stochastic stability of the closed-loop system.
By analyzing the iteration algorithm's dynamic behavior, we demonstrated that the VoI metric exhibits symmetry and, when the system matrix is diagonalizable, is monotone and quasi-convex. Building on these findings, we established that the optimal scheduling law exhibits a threshold structure and takes a quadratic form. The region for the threshold value was also characterized. 

\section{Appendix}
\begin{definition}\label{definition:equicontinuous}(\textbf{Pointwise equicontinuity}\cite{narici2011topological}) Let $\HC$ be a family of real-valued functions on a metric space $\XC$,
$\HC$ is said to be equicontinuous at the point $x \in \XC$ if for every $r > 0$, there exists a $\epsilon > 0 $ such that $d(h(x), h(y)) < r$ for all $h \in \HC$ and all $y$ such that $d(x, y) < \epsilon$.  The family $\HC$ is pointwise equicontinuous if it is equicontinuous at each point of $\XC$. 
\end{definition}
\textit{Proof of Condition 6) in Proposition~\ref{proposition:bellman}.} From Proposition~\ref{proposition:bellman}, we have $\vert h_{\alpha_n}(x)- h_{\alpha_n}(y)\vert = | J_{\alpha_n}^\ast(x)-J_{\alpha_n}^\ast(y) |  $, for $x,y\in \XC$. Moreover, $J_{\alpha_n}^\ast$ satisfies the dynamic programming equation \cite{hernandez1992discrete}, i.e., $    J_{\alpha_n}^\ast = \min_{\delta \in \AC(e)} \big\{ g(e,\delta) + \alpha_n \int J_{\alpha_n}^\ast(y) \PC({\rm d}y \vert e, \delta) \big\}.$ 
Denote $k := (e,\delta)$ and  $k' := (e',\delta')\in \KC$. Denote $d_1(e,e'):= \|e-e'\|$ and $d_2(\delta,\delta'):=|\delta -\delta'|$ as metrics on $\XC$ and $\AC$, respectively. For all $k,k'\in \KC$, denote $d(k,k'):= \max\big\{d_1(e,e'),d_2(\delta,\delta') \big\}$ as the metric on $\KC$. 
Moreover, let $\psi$ be the class of nondecreasing functions $\psi : [0,\infty) \rightarrow [0,\infty)$ such that $\lim_{s\downarrow0} \Phi(s) = 0$. 
If there exist functions $\psi_e^g,\psi_e^p \in \Phi$ such that
\begin{enumerate}
    \item $\big|g_{\alpha_n}(k)-g_{\alpha_n}(k')\big| \le \psi_e^g\big(d(k,k')\big) $; 
    \item $  \big\vert \int J_{\alpha_n}^\ast(y)  \PC( {\rm d}y \big\vert k) -  \int J_{\alpha_n}^\ast(y) \PC({\rm d}y \vert k')\vert   \le \psi_e^p\big(d(k,k')\big) $, 
\end{enumerate}
for all $k, k' \in \KC$, then $h_{\alpha_n}$ is pointwise equicontinuous. 
Regarding the left-hand side of the first inequality, we have 
\begin{eqnarray*}
 &&  \big|g_{\alpha_n}(k)-g_{\alpha_n}(k')\big| \nonumber \\ 
 \eq \big|\delta\theta + (1-\delta)(Ae)^\top  \Sigma A e - \delta'\theta - (1-\delta')(Ae')^\top \Sigma A e'\big| \nonumber \\ 
 \qle \theta|\delta-\delta'| + \big|(1-\delta)(Ae)^\top \Sigma A e  - (1-\delta')(Ae)^\top \Sigma A e \nonumber \\ 
 && + (1-\delta')(Ae)^\top \Sigma A e - (1-\delta')(Ae')^\top \Sigma A e'\big| \nonumber \\ 
\qle \big(\theta + (Ae)^\top \Sigma A e\big)\big|\delta-\delta'\big|  \nonumber \\
&& + \big|1-\delta'\big| \bar{\rho}(A^\top \Sigma A) \big\| e + e'\big\| \big\|e - e'\big\| \nonumber \\ 
\qle  b_1 d(k,k')
\end{eqnarray*} 
with $b_1 =|1-\delta'| \bar{\rho}(A^\top \Sigma A) \| e + e'\|   +  \theta + (Ae)^\top \Sigma A e $. 
Thus, inequality 1) holds.   

Condition 4) of Proposition~\ref{proposition:bellman} shows that $\{J_{\alpha_n}^\ast\}$, for all $n\ge0$, is a nonnegative and bounded function sequence. Then we denote its upperbound as $\bar{J}_{\alpha_n}^\ast$.  
The left-hand side of the inequality 2) is written as 
\begin{eqnarray}\label{eq:J k k' bound 1}
&& \left\vert\int J_{\alpha_n}^\ast(y) \PC({\rm d}y \vert k) - J_{\alpha_n}^\ast(y)  \PC({\rm d}y \vert k') \right\vert \nonumber \\ 
\qle \bar{J}_{\alpha_n}^\ast  \left\vert \int \PC({\rm d}y \vert k) - \PC({\rm d}y \vert k')   \right\vert \nonumber \\ 
\qle \bar{J}_{\alpha_n}^\ast \int\big|p(y|k) - p(y|k')\big| {\rm d}y.  
% \qle  \bar{V}_{\alpha_n} \int| \delta p_{\xi}(y)   + (1-\delta ) p_{\xi}(y-Ae ) \nonumber \\ 
% && - \delta' p_{\xi}(y)   + (1-\delta' ) p_{\xi}(y-Ae' ) | {\rm d}y. 
\end{eqnarray}
Moreover, we have 
\begin{eqnarray}\label{eq:prob bound 1}
&& \big|p(y|k) - p(y|k')\big| \nonumber \\
\qle    \big| \delta p_{\xi}(y)   + (1-\delta ) p_{\xi}(y-Ae ) \nonumber \\ 
&& - \delta' p_{\xi}(y) - (1-\delta' ) p_{\xi}(y-Ae' ) \big|   \nonumber \\ 
\qle \big|\delta-\delta'\big|p_{\xi}(y) + \big|(1-\delta ) p_{\xi}(y-Ae ) + (1-\delta' ) p_{\xi}(y-Ae ) \nonumber \\ 
&& - (1-\delta' ) p_{\xi}(y-Ae ) - (1-\delta' ) p_{\xi}(y-Ae' )\big| \nonumber \\
\qle \big|\delta-\delta'\big|\big(p_\xi (y ) +p_{\xi}(y-Ae)\big) \nonumber \\
&& + \big|1-\delta'\big|\big|p_{\xi}(y-Ae) - p_{\xi}(y-Ae')\big|. 
\end{eqnarray}
Let $z: = y-Ae$, we have  
\begin{align}\label{eq:prob bound 2}
&    \int \big|p_{\xi}(y-Ae) - p_{\xi}(y-Ae')\big| {\rm d}y \nonumber \\
\hspace{-2em}& =\hspace{-0.5em}  \int\big |p_{\xi}(z) - p_{\xi}\big(z+A(e-e')\big)\big|{\rm d}z \le \hspace{-0.5em} \int L_0\|e-e'\| {\rm d}z
\end{align} 
with $L_0 = \max_{z \in \XC} \|\nabla p_\xi (z)\| $ being a bounded positive scalar, where the boundedness is from the definition of $p_\xi$. 
% The last inequality establishes as $p_{\xi}(z)$ is Lipschtiz continuous in $z \in \XC$, and then $ \|\nabla p_\xi (z)\|$ is bounded in $z$. 
Substitute \eqref{eq:prob bound 1} and \eqref{eq:prob bound 2} into \eqref{eq:J k k' bound 1}, we have 
\begin{eqnarray*}
&& \left\vert\int J_{\alpha_n}^\ast(y) \PC({\rm d}y \vert k) - J_{\alpha_n}^\ast(y)  \PC({\rm d}y \vert k') \right\vert      \nonumber \\
\qle \bar{J}_{\alpha_n} \int \big(p_\xi(y) + p_{\xi}(y-Ae)\big)d_1(\delta,\delta')\nonumber \\
&& + |1-\delta'|L_0 d_2(e,e')  {\rm d}y  \le \bar{J}_{\alpha_n} \int  b_2 d(k,k') {\rm d}y 
\end{eqnarray*}
with $b_2 = \big(p_\xi(y) + p_{\xi}(y-Ae)\big) + |1-\delta'|L_0$.
Thus, inequality 2) holds. Until here, we show that the family of $J_{\alpha_n}^\ast$ are pointwise equicontinuous. Since $\vert h_{\alpha_n}(x)- h_{\alpha_n}(y)\vert = | J_{\alpha_n}^\ast(x)-J_{\alpha_n}^\ast(y) |  $, $h_{\alpha_n}$ are pointwise equicontinuous.
\hfill $\blacksquare$

\noindent\textit{Proof of Lemma~\ref{lemma:monotonicity Eh}}. 
As proved in Proposition~\ref{proposition:he symmetry} that $\Eb\big[J_t^\ast(\Lambda s + \zeta)\big]$ is symmetric in $s$, 
we first discuss the monotonicity of $\Eb\big[J_t^\ast(\Lambda s + \zeta)\big]$ 
when $s_i\ge 0$, for $i\in \NC$ and then extend this result to the case $s_i < 0$. 
In the following, we will show that, for $i \in \NC$, if $J_t^\ast(\vec{\epsilon}_i) \ge J_t^\ast (\vec{s}_i)$ for $\epsilon_i>s_i>0$, then $\Eb\big[J_t^\ast(\Lambda\vec{\epsilon}_i+\zeta)\big] \ge \Eb\big[J_t^\ast(\Lambda\vec{s}_i+\zeta)\big] $. Denote $\Lambda_{ij}$ as the $i$-th row and $j$-th column component of matrix $\Lambda$.
Define a region $\SC_1^- := \big\{\zeta \in \XC \big| \zeta_i \le -  \Lambda_{ii}(\vec{s}_i+\vec{\epsilon}_i)/2  \big\}$ and $\SC_1^+ := \XC \backslash \SC_1^-$. 
Then we have 
\begin{eqnarray}\label{eq:J_k mono}
&&\Eb\big[J_t^\ast(\Lambda\vec{\epsilon}_i+\zeta)\big] - \Eb\big[J_t^\ast(\Lambda\vec{s}_i+\zeta)\big] \nonumber \\
    % \eq \int_{\XC
    % } \big(J_t^\ast(\Lambda\vec{\epsilon}_i +\zeta)-J_t^\ast(\Lambda\vec{s}_i +\zeta) \big)p_\zeta(\zeta) {\rm d}\zeta  \nonumber \\
    \eq \int_{\SC_1^-}   \big(J_t^\ast(\Lambda\vec{\epsilon}_i +\zeta)-J_t^\ast(\Lambda\vec{s}_i +\zeta) \big)p_\zeta(\zeta) {\rm d}\zeta \nonumber \\ 
    && +\int_{\SC_1^+}  \big(J_t^\ast(\Lambda\vec{\epsilon}_i +\zeta)-J_t^\ast(\Lambda\vec{s}_i +\zeta) \big)p_\zeta(\zeta){\rm d}\zeta \nonumber \\ 
     \eq \int_{\SC_1^-}   \big(J_t^\ast(-\Lambda\vec{\epsilon}_i -\zeta)-J_t^\ast(-\Lambda\vec{s}_i -\zeta) \big)p_\zeta(\zeta) {\rm d}\zeta \nonumber \\ 
    && +\int_{\SC_1^+}  \big(J_t^\ast(\Lambda\vec{\epsilon}_i +\zeta)-J_t^\ast(\Lambda\vec{s}_i +\zeta) \big)p_\zeta(\zeta){\rm d}\zeta, 
\end{eqnarray}
where the last equality establishes as $J_t^\ast(e) =J_t^\ast(-e) $. 
Define a region $\SC_2^+ := \big\{\zeta \in \XC \big| \zeta_i \ge   \Lambda_{ii}(\vec{s}_i+\vec{\epsilon}_i)/2\big\}$.
Let $z :=-\zeta$ and $y:=\zeta+\Lambda\vec{s}_i+\Lambda\vec{\epsilon}_i$, \eqref{eq:J_k mono} is further written as 
\begin{eqnarray*}
    \eq \int_{\SC_2^+} 
      \big(J_t^\ast(-\Lambda\vec{\epsilon}_i +z)-J_t^\ast(-\Lambda\vec{s}_i 
     +z) \big)p_\zeta(z) {\rm d}z \nonumber \\
    &&\hspace{-1em} +\int_{\SC_2^+}  \big(J_t^\ast(y-\Lambda\vec{s}_i)-J_t^\ast(y-\Lambda\vec{\epsilon}_i) \big)p_\zeta(y-\Lambda\vec{\epsilon}_i-\Lambda\vec{s}_i){\rm d}y \nonumber \\ 
    %  \eq \int_{\SC_2^+} 
    %   \big(J_t^\ast(z-\Lambda\vec{\epsilon}_i )-J_t^\ast(z-\Lambda\vec{s}_i) \big)p_\zeta(z) {\rm d}z\nonumber \\ 
    % &&\hspace{-1em} +\int_{\SC_2^+}  \big(J_t^\ast(z-\Lambda\vec{s}_i)-J_t^\ast(z-\Lambda\vec{\epsilon}_i) \big)p_\zeta(z-\Lambda\vec{\epsilon}_i-\Lambda\vec{s}_i){\rm d}z \nonumber \\ 
    \eq  \int_{\SC_2^+} \big( J_t^\ast(z-\Lambda\vec{s}_i)-J_t^\ast(z-\Lambda\vec{\epsilon}_i)\big) \nonumber \\
    &&\big( p_\zeta(z-\Lambda\vec{s}_i-\Lambda\vec{\epsilon}_i) - p_\zeta (z)\big) {\rm d}z,
\end{eqnarray*}
where the first equality is from $p_\zeta(z) = p_\zeta (-z)$. 
Denote $z_i$ as the $i$-th component of $z$. 
For $i \in \NC$ and $\epsilon_i > s_i>0$, 
\begin{enumerate}
    \item when $\Lambda_{ii} \ge 0$ and $z_i \ge \Lambda_{ii}\epsilon_i $, 
we have $z_i-\Lambda_{ii}s_i   \ge  z_i-\Lambda_{ii}\epsilon_i $ and $  z_i-\Lambda_{ii}(s_i+\epsilon_i)  \le  z_i $;
\item when $\Lambda_{ii} \ge 0$ and $z_i \in \big[ \Lambda_{ii}(s_i+\epsilon_i)/2,\Lambda_{ii}\epsilon_i\big] $, we have $| z_i-\Lambda_{ii}s_i | \ge | z_i-\Lambda_{ii}\epsilon_i |$ and $| z_i-\Lambda_{ii}(s_i+\epsilon_i) | \le |z_i|$; 
\item when $\Lambda_{ii} < 0$ and  $z_i \ge \Lambda_{ii}\epsilon_i $, 
we have $ z_i-\Lambda_{ii}s_i   \le   z_i-\Lambda\epsilon_i  $ and $ z_i-\Lambda_{ii}(s_i+\epsilon_i)   \ge  z_i $;
\item when  $\Lambda_{ii} < 0$ and $z_i \in \big[\Lambda_{ii}(s_i+\epsilon_i)/2, \Lambda_{ii}\epsilon_i \big] $, we have $| z_i-\Lambda_{ii}s_i | \ge | z_i-\Lambda_{ii}\epsilon_i |$ and $| z_i-\Lambda_{ii}(s_i+\epsilon_i)| \le |z_i|$.
\end{enumerate}
Note that $J_t^\ast(z)$ and $p_\zeta(z)$ are symmetric functions in $z\in\XC$, $J_t^\ast(z)$  monotonically nondecreases with the increasing $|z_i|$ and $p_\zeta(z)$  monotonically decreases with the increasing $|z_i|$.  
For all the above cases, we have $\big( J_t^\ast(z-\Lambda\vec{s}_i)-J_t^\ast(z-\Lambda\vec{\epsilon}_i)\big) \big( p_\zeta(z-\Lambda\vec{s}_i-\Lambda\vec{\epsilon}_i) - p_\zeta (z)\big) \ge 0$. Thus, we obtain that $\Eb\big[J_t^\ast(\Lambda\vec{\epsilon}_i+\zeta)\big] \ge \Eb\big[J_t^\ast(\Lambda\vec{s}_i+\zeta)\big] $ establishes for $\epsilon_i>s_i\ge0$ and $i \in \NC$.  
Following similar arguments, $\Eb\big[J_t^\ast(\Lambda\vec{\epsilon}_i+\zeta)\big] \ge \Eb\big[J_t^\ast(\Lambda\vec{s}_i+\zeta)\big] $ 
holds for $\epsilon_i<s_i\le0$ and $i \in \NC$. 

To sum up, for $i \in \NC$, if $J_t^\ast(s)$ monotonically nondecreasing in $s_i$ for $s_i \ge 0$, and monotonically nonincreasing in $s_i$ for $s_i < 0$, then so as $\Eb\big[J_t^\ast(\Lambda s+\zeta)\big]$. 
\hfill $\blacksquare$

% \section*{Acknowledgment}
% We thank Dr. Yuchao Li and Mr. Zifan Wang for their helpful discussion and feedback.
 
% \section*{References and Footnotes}

% \section{Submitting Your Brief for Review}
% \section{Publication Principles}

\bibliographystyle{ieeetr}
\bibliography{references}

\vspace{-2em}

\end{document}